\documentclass{fac}

\PassOptionsToPackage{prologue}{xcolor}

\newif\ifarxiv
\arxivtrue

\usepackage{booktabs}
\usepackage{subcaption}
\usepackage{amssymb}
\usepackage{amsmath}
\usepackage[dvipsnames]{xcolor}
\usepackage{multirow}
\usepackage[linesnumbered]{algorithm2e}
\usepackage{multicol}
\usepackage{mdframed}
\usepackage{listings}
\usepackage{hyperref}
\usepackage{tikz}
\usepackage{todonotes}

\newtheorem{definition}{Definition}[section]

\usetikzlibrary{shapes.geometric, arrows}

\definecolor[named]{lipicsLightGray}{rgb}{0.85,0.85,0.86}

\lstdefinelanguage{julia}
{
  keywordsprefix=\$,
  morekeywords={
    exit,whos,edit,load,is,isa,isequal,typeof,tuple,ntuple,uid,hash,finalizer,convert,promote,
    subtype,typemin,typemax,realmin,realmax,sizeof,eps,promote_type,method_exists,applicable,
    invoke,dlopen,dlsym,system,error,throw,assert,new,Inf,Nan,pi,im,begin,while,for,in,return,
    break,continue,macro,quote,let,if,elseif,else,try,catch,end,bitstype,ccall,do,using,module,
    import,export,importall,baremodule,immutable,local,global,const,Bool,Int,Int8,Int16,Int32,
    Int64,Uint,Uint8,Uint16,Uint32,Uint64,Float32,Float64,Complex64,Complex128,Any,Nothing,None,
    function,type,typealias,abstract
  },
  sensitive=true,
  morecomment=[l]{\#},
  morestring=[b]',
  morestring=[b]" 
}

\newcommand{\K}{\mathbb{K}}
\newcommand{\Q}{\mathbb{Q}}
\newcommand{\N}{\mathbb{N}}

\newcommand{\CC}{\mathcal{C}} 

\renewcommand{\vec}[1]{\boldsymbol{#1}}

\newcommand{\seq}[1]{\left({#1}\right)_{n=0}^\infty}

\newcommand{\kth}{\textsf{th}}
\newcommand{\ini}[1]{\bar{#1}}

\newcommand{\geom}{\omega}

\newcommand{\polysin}{\sqsubset}

\newcommand{\reserved}{\mathbf}
\newcommand{\WHILE}{\reserved{while}}
\newcommand{\DO}{\reserved{do}}
\newcommand{\END}{\reserved{end}}

\newcommand{\tool}[1]{\texttt{#1}}

\DeclareMathOperator{\cstr}{cstr}
\DeclareMathOperator{\decompose}{split}

\newcommand{\absynth}{\tool{Absynth}}
\newcommand{\absynthurl}{\url{https://github.com/ahumenberger/Absynth.jl}}

\newcommand{\abbrv}[1]{{#1}}

\binoppenalty=\maxdimen
\relpenalty=\maxdimen

\newtheorem{theorem}{Theorem}[section]
\newtheorem{proposition}{Proposition}[section]
\newtheorem{lemma}{Lemma}[section]

\newtheorem{example}{Example}[section]
\newtheorem{remark}{Remark}[section]

\newcommand{\revision}[1]{#1}

\newcommand{\qed}{}

\title[Algebra-Based Reasoning for Loop Synthesis]
      {Algebra-Based Reasoning for Loop Synthesis}

\author[Humenberger et al]
    {Andreas Humenberger$^1$, Daneshvar Amrollahi$^1$, Nikolaj Bj\o{}rner$^2$ and Laura Kov\'acs$^1$ \\
     $^1$TU Wien, Austria\\
     $^2$Microsoft Research}

\correspond{Laura Ko\'acs, TU Wien,              e-mail: laura.kovacs@tuwien.ac.at}

\pubyear{2000}
\pagerange{\pageref{firstpage}--\pageref{lastpage}}

\begin{document}

 \makecorrespond

\maketitle

\begin{abstract}
Provably correct software is one of the key challenges of our software-driven
society. Program synthesis -- the task of constructing a program satisfying a
given specification -- is one strategy for achieving this. The result of this
task is then a program which is correct by design. As in the domain of program
verification, handling loops is one of the main ingredients to a successful
synthesis procedure.

We present an algorithm for synthesizing loops satisfying a given polynomial
loop invariant. The class of loops we are considering can be modeled by a system
of algebraic recurrence equations with constant coefficients, encoding
thus program loops with affine operations among program variables. We turn the
task of loop synthesis into a polynomial constraint problem, by precisely
characterizing the set of all loops satisfying the given
invariant. We prove
 soundness of our approach, as well as its completeness with respect to an a
 priori fixed upper bound on the number of program variables.
  Our work has 
  applications towards synthesizing loops 
  satisfying a given polynomial loop invariant, program verification, 
  as well as generating number sequences from algebraic relations. 
  To understand viability of the methodology and heuristics for synthesizing loops, we implement and evaluate 
  the method using the \absynth{} tool.

  \bigskip
\noindent  This paper is an extended version of the ``Algebra-Based Loop
  Synthesis'' manuscript published
at iFM 2020. 
\end{abstract}


\section{Introduction}\label{sec:intro}

The two most rigorous approaches for providing correct software are
given by formal program verification and program synthesis~\cite{GulwaniPOPL10}. The task
of formal verification is to prove correctness of a given program with respect
to a given (logical) specification. On the other hand, program
synthesis aims at
generating programs which adhere to a given specification. The result of a
synthesis problem is therefore a program which is correct by construction. While
formal verification has received considerable attention with impressive results,
for example, in ensuring safety of device drivers~\cite{SLAM11} and
security of 
web services~\cite{Cook18}, program synthesis turns out to be an algorithmically
much more difficult challenge~\cite{KuncakACM12}.

The classical setting of
program synthesis has been to synthesize programs from proofs of logical
specifications that relate the inputs and the outputs, in short inputs-outputs,
of the program~\cite{MannaW80}.
Thanks to recent successful trends in formal verification based on automated
reasoning~\cite{Z3,Yices,vampire13}, this traditional view of program synthesis
has been refined to the setting of syntax-guided synthesis
(SyGuS)~\cite{Alur15}.   In addition to logical
specifications, SyGuS approaches consider further constraints on the program
templates to be synthesized, thus limiting the search space of possible
solutions.
A wide range of efficient
applications of SyGuS have so far emerged, for example programming by
examples~\cite{GulwaniIJCAR16}, component-based synthesis~\cite{GulwaniICSE10}
with learning~\cite{DilligPLDI18} and sketching~\cite{SolarICML19}.

Yet, both in the setting of verification and synthesis, one of the main
challenges is to verify and/or synthesize loops/recursion. In verification,
solving this challenge requires generating additional program assertions in the
form of \emph{loop
invariants}~\cite{Rodriguez-CarbonellK07,HumenbergerJK17,KincaidCBR18}.
Intuitively, a loop invariant is a formal description of the behavior of the
loop, expressing loop properties that hold before and after each loop iteration.
In synthesis, the challenge of reasoning with loops comes by answering the
question whether there exists a loop satisfying a given loop invariant and
constructively synthesizing a loop with respect to a given invariant. We refer
to this task of synthesis as {\it loop synthesis}. Note that loop synthesis can
be considered as the reverse problem of loop invariant generation: rather than
generating invariants summarizing a given loop, we synthesize loops whose
summaries are captured by a given invariant property.

In this paper, we present an algorithmic approach to loop
synthesis, by relying on the theory
of algebraic recurrence equations~\cite{KauersP11}. We synthesize and optimize loops whose
functional behavior is captured by a given invariant ${p(\vec{x})=0}$, such
that the synthesized loops use only affine computations among $\vec{x}$. 
We define {\it our loop synthesis task} as follows:

\vspace*{1.0em}
\begin{mdframed}[frametitle=Loop Synthesis, 
      frametitlefont=\small\sffamily\MakeUppercase,
      innertopmargin=3pt,
      frametitleaboveskip=9pt,
      innerbottommargin=8pt
    ]
  \textbf{Given a polynomial} property $p(\vec{x})$ over a set $\vec{x}$ of variables,\\
  \textbf{generate a loop} $\mathcal{L}$ with affine operations over  variables
  $\vec{x}$,  such that
  $p(\vec{x})=0$ is an 
    invariant of the loop.
\end{mdframed}
\vspace*{1.0em}

While our  loop synthesis problem is formulated such that only one
loop satisfying a given polynomial invariant $p(\vec{x})=0$ is
captured, we note that 
in our work we first characterize \emph{all} loops for which
 $p(\vec{x})=0$ is an invariant (Section~\ref{sec:synth}) and then
 extract one loop by optimizing the set of all loop solutions
 (Section~\ref{sec:automate}). One such optimization criteria comes, 
 for example, with avoiding trivial solutions, that is avoiding loops
 whose affine updates do not change values of loop variables.
 

We believe our work  is the first complete technique
for synthesizing loops from (non-linear) polynomial
invariants. The \emph{key ingredient} of our work comes with the
reduction of loop synthesis to the problem of solving algebraic recurrences
among loop variables. The  \emph{inner algorithmic magic} of our
approach is an SMT-based solution  towards solving non-linear
algebraic constraints, 
allowing us to test existential properties of bounded degree
polynomials (invariants) to derive
universal (loop) relations.

\paragraph{Motivating Example.} 
Let us first motivate loop synthesis using Figure~\ref{fig:Dafny:a}. The loop is
based on an online tutorial\footnote{\url{https://rise4fun.com/Dafny/}} of the
Dafny verification framework~\cite{Dafny17}: Figure~\ref{fig:Dafny:a} uses only
affine updates among its variables, and the task is to revise/repair
Figure~\ref{fig:Dafny:a} into a partially correct program with respect to the
precondition $N\geq 0$ and post-condition $c=N^3$ such that the polynomial loop
invariant $n \leq N \wedge c=n^3 \wedge k=3n^2+3n+1 \wedge m=6n+6$ holds. \revision{We note that our focus in on partially correct programs~\cite{Hoare69}, and thus we do not consider total correctness, that is loop termination. }


\begin{figure}
  \begin{subfigure}{.31\textwidth}
    \begin{center}
      \begin{tabular}{l}
      $(c, k, m, n) \gets (0, 0, 0, 0)$\\
      $\WHILE~n<N~\DO$\\
      \quad$c \gets c+k$\\
      \quad$k \gets k+m$\\
      \quad$m \gets m+9$\\
      \quad$n \gets n+1$\\
      $\END$
      \end{tabular}
    \end{center}
  \caption{Faulty loop}\label{fig:Dafny:a}
  \end{subfigure}\hfill
  \begin{subfigure}{.34\textwidth}
    \begin{center}
      \begin{tabular}{l}
      $(c, k, m, n) \gets (0, \colorbox{lipicsLightGray}{$1$}, \colorbox{lipicsLightGray}{$6$}, 0)$\\
      $\WHILE~n<N~\DO$\\
      \quad$c \gets c+k$\\
      \quad$k \gets k+m$\\
      \quad$m \gets m+\colorbox{lipicsLightGray}{$6$}$\\
      \quad$n \gets n+1$\\
      $\END$
      \end{tabular}
    \end{center}
  \caption{Synthesized loop}\label{fig:Dafny:b}
  \end{subfigure}\hfill
  \begin{subfigure}{.34\textwidth}
    \begin{center}
      \begin{tabular}{l}
      $(c, k, m, n) \gets (0, \colorbox{lipicsLightGray}{$1$}, \colorbox{lipicsLightGray}{$6$}, 0)$\\
      $\WHILE~n<N~\DO$\\
      \quad$c \gets c+k$\\
      \quad$k \gets k+\colorbox{lipicsLightGray}{$6n + 6$}$\\
      \quad$m \gets m+\colorbox{lipicsLightGray}{$6$}$\\
      \quad$n \gets n+1$\\
      $\END$
      \end{tabular}
    \end{center}
  \caption{Synthesized loop}\label{fig:Dafny:c}
  \end{subfigure}
  \caption{
Examples of loop    synthesis. 
    Figures~\ref{fig:Dafny:b}-\ref{fig:Dafny:c} are
    revised versions of Figure~\ref{fig:Dafny:a} such that the expression
    ${c=n^3 \wedge k=3n^2+3n+1 \wedge m=6n+6}$ is an invariant of Figures~\ref{fig:Dafny:b}-\ref{fig:Dafny:c}.}
  \label{fig:Dafny}
\end{figure}

In this paper we introduce an algorithmic approach to loop synthesis by relying on
algebraic recurrence equations and constraint solving over polynomials. In
particular,  we automatically synthesize
Figures~\ref{fig:Dafny:b}-\ref{fig:Dafny:c} by using the given non-linear
polynomial equalities $c=n^3 \wedge k=3n^2+3n+1 \wedge m=6n+6$ as input
invariant to our loop synthesis task. Both synthesized programs, with the
loop guard ${n<N}$ as in Figure~\ref{fig:Dafny:a}, are partially correct
program with respect to the given requirements. Moreover,
Figures~\ref{fig:Dafny:b}-\ref{fig:Dafny:c} precisely capture the
solution space of  $c=n^3 \wedge k=3n^2+3n+1 \wedge m=6n+6$, by
implementing only affine operations. 

\paragraph{Algebra-based Loop Synthesis.} 
Inspired by syntax-guided synthesis -- SyGuS~\cite{Alur15}, we consider our loop synthesis task  with
additional requirements on the loop to be synthesized: we  impose
syntactic  requirements on the form of loop expressions and
guards to be synthesized. The imposed requirements allow us to 
\emph{reduce the loop synthesis task to
the problem of generating linear/affine recurrences with constant
coefficients, called C-finite recurrences~\cite{KauersP11}}.
As such, \emph{loop synthesis} provides an algorithmic
solution to the following loop reasoning challenge:
\emph{Given a polynomial $p(\vec{x})$ over loop
variables $\vec{x}$, how can the entire solution
space of ${p(\vec{x})=0}$ be iteratively computed using only affine
operations inducing C-finite number sequences among $\vec{x}$?}




Our approach to synthesis is however 
conceptually different from other SyGuS-based methods, such
as~\cite{GulwaniIJCAR16,DilligPLDI18,SolarICML19}: rather than iteratively
refining both the input and the solution space of synthesized programs, we take
polynomial relations describing a potentially infinite set of input values and
precisely capture not just one loop, but the \emph{set of all loops} (i)~whose invariant is given by
our input polynomial and (ii)~whose variables induce C-finite number
sequences.
Any instance of this set therefore yields a loop that is
partially correct by construction and only implements affine computations. Figures~\ref{fig:Dafny:b}-\ref{fig:Dafny:c} depict two solutions of our loop synthesis task for the
invariant ${c=n^3 \wedge k=3n^2+3n+1 \wedge m=6n+6}$.



The main steps of our approach are as follows. (i)~Let $p(\vec{x})$ be a polynomial
over variables $\vec{x}$ and let ${s\geq 0}$ be an upper bound on the number of
program variables to be used  in the loop. If not specified, $s$ is considered to
be the number of variables from $\vec{x}$. 
%
%
%
(ii)~We use syntactic constraints over the loop body to be synthesized and
define a loop template, as given by our programming
model~\eqref{eq:loop}. Our programming model
imposes that 
 the functional behavior of the synthesized loops can
be modeled by a system of C-finite recurrences (Section~\ref{sec:model}). 
(iii)~By using the invariant property of $p(x)=0$ for the loops to the
synthesized, we construct a polynomial
constraint problem (PCP) characterizing the set of all
loops satisfying the constraints of~\eqref{eq:loop} for which ${p(x) = 0}$
is a loop invariant (Section~\ref{sec:synth}). 
Our approach combines symbolic computation techniques over algebraic recurrence equations
with polynomial constraint solving. We prove that our approach to loop synthesis
is both \emph{sound} and \emph{complete}. By completeness we mean that if
there is a loop $\mathcal{L}$ with at most $s$ variables satisfying the
invariant ${p(\vec{x})=0}$ such that the loop body meets our C-finite/affine syntactic
requirements, then $\mathcal{L}$ is synthesized by our method (Theorem~\ref{thm:sound-complete}).
Moving beyond $s$, that is, deriving an upper bound on
the number of program variables from the invariant, is interesting
further work, 
with connections to the {inverse problem of
difference Galois theory}~\cite{Galois}.


We finally note that our work is
not restricted to specifications given by a single polynomial equality
invariant. Rather, the invariant given as input to our synthesis approach
can be conjunctions of polynomial equalities -- as also shown in
Figure~\ref{fig:Dafny}.

\paragraph{Beyond Loop Synthesis.}
Our work has applications beyond loop synthesis -- such as in
generating number sequences from algebraic relations and program/compiler optimizations.
\begin{itemize}
\item\label{usecase:seq}
\emph{Generating number sequences.} Our approach provides a partial solution to
an open mathematical problem: given a polynomial relation among number
sequences, e.g.
%
\begin{equation}
  \label{eq:fib:relation}
  f(n)^4 + 2f(n)^3f(n+1) - f(n)^2f(n+1)^2 - 2f(n)f(n+1)^3 + f(n+1)^4 = 1,
\end{equation}
synthesize algebraic recurrences defining these sequences. There exists no
complete method for solving this challenge, but we give a complete approach in
the C-finite setting parameterized by an a priori bound $s$ on the order of the
recurrences. For the given relation~\eqref{eq:fib:relation} among $f(n)$ and $f(n+1)$, our
work generates the C-finite recurrence equation $f(n+2)=f(n+1)+f(n)$ which
induces the Fibonacci sequence.

\item\label{usecase:opt}
\emph{Program optimizations.} Given a polynomial invariant, our approach
generates a PCP such that any solution to this PCP yields a loop
satisfying the given invariant. 
By using additional constraints
encoding a cost function on the loops to be synthesized, our method can be
extended to synthesize loops that are optimal with respect to the considered
costs, for example  synthesizing loops that use only addition in
variable updates. 
Consider for example Figures~\ref{fig:Dafny:b}-\ref{fig:Dafny:c}: 
the loop body of Figure~\ref{fig:Dafny:b} uses only addition, whereas
Figure~\ref{fig:Dafny:c} implements also multiplications by
constants. 


\item\label{usecase:compiler}
  \emph{Compiler optimizations.}
To reduce execution time spent within loops, 
compiler optimization techniques, such as strength reduction,
aim at replacing expensive loop operations with semantically
equivalent but less expensive operations~\cite{SRed01}.
One such optimization within strength reduction replaces
``strong" loop multiplications by additions among program
variables. The burden of strength reductions comes however with 
identifying inductive loop variables and invariants to
be used for loop optimization. Our loop synthesis method 
therefore serves as a foundation for strength reduction
  optimization, by converting polynomial (loop) expressions into
  incremental affine (loop) computations.

\item\label{usecase:invgen}
\emph{Invariant generation \revision{and loop equivalence}.} Our approach may improve the state-of-the-art in \revision{polynomial} 
invariant generation, as follows.  By considering a loop $L$, let $I$ denote the
\revision{polynomial} loop invariant of $L$ generated by existing invariant generation
approaches, 
such as~\cite{Rodriguez-CarbonellK07,HumenbergerJK17,KincaidCBR18,Worrell18}. By
using $I$ as our input,  
our technique synthesizes a loop $L'$ that is
(relational) equivalent to $L$ in the sense that both $L$ and $L'$ have the same
invariant $I$. \revision{We showcase such use case of our synthesis approach in Section~\ref{sec:beyond:equiv}}. We further note, that our loop synthesis task can also be used to check
soundness/completeness  of existing invariant generation techniques: apply
invariant generation approaches on $L'$ with the goal of deriving the invariant $I$. 

%

\item\label{usecase:teach}
\emph{Teaching formal methods.} Finally, our work can also be used in solving
formal verification challenges, as follows. By 
considering examples that are not partially correct with respect to a given
invariant, the task is to revise these programs into correct ones. By
using the given invariants as inputs to our approach, we automatically infer partially correct
programs, thus allowing
users/teachers/practitioners of formal methods to repair
programs in a fully automated manner. This use-case of loop synthesis 
has been deployed in our ``Formal Methods in Computer Science'' master
course at the TU Wien, in particular in generating and correcting
course assignments on deductive program verification, \revision{as discussed in Section~\ref{sec:beyond:teach}}.

\end{itemize}

\paragraph{Contributions.}
This paper brings integrated approaches to formal modelling and
analysis of software, by combining symbolic computation, program
analysis and SMT reasoning. In summary, we make the following
contributions. 
\begin{itemize}
\item
We propose an \emph{automated procedure for synthesizing loops} that are
partially correct with respect to a given polynomial loop invariant
(Section~\ref{sec:synth}). 
By exploiting properties of C-finite sequences, we construct a PCP which
precisely captures \emph{all solutions} of our loop synthesis task. 
We are not aware of previous approaches synthesizing loops from (non-linear) polynomial
invariants. 

\item 
We also synthesize the initial values of the loop variables, that is, the values
before executing the loop. We first consider loops with concrete initial values,
so-called \emph{non-parameterized} loops (Section~\ref{sec:synth:nonparam}).
We then refine our technique towards the synthesis of
\emph{parameterized} loops, that is, loops with symbolic initial values
(Section~\ref{sec:synth:param}). 


\item
We prove that our approach to loop synthesis is sound and complete
(Theorem~\ref{thm:sound-complete}). That is, if there is a loop whose invariant
is captured by our given specification, our approach synthesizes this loop. To
this end, we consider completeness modulo an a priori fixed upper bound $s$ on
the number of loop variables.

\item
  We extend our task of loop synthesis with additional constraints,
  for optimizing the solution space of our PCP (Section~\ref{sec:automate}).
These optimizations are essential in automating loop synthesis. 

%
\item 
We implemented our approach in the new open-source framework \absynth{}. We first evaluated our work on a number of
academic examples on loop analysis as well as on generating number
sequences in algorithmic combinatorics (Section~\ref{sec:experiments}). \

\item \revision{We further used our work in the context of loop equivalence and  report on our experience in using loop synthesis as a work-horse for teaching formal methods at the TU Wien (Section~\ref{sec:beyond}).} 
\end{itemize}

\paragraph{Relation to our previous work~\cite{HumenbergerBK20}.} This paper extends our previous work~\cite{HumenbergerBK20}  in a number of ways, as summarized below.
In Section~\ref{sec:prelim}, we provide additional algebraic results
and insights needed for translating the task of loop into  a
recurrence solving problem in Section~\ref{sec:synth:param}.  Soundness and
completeness proofs of our key results are given in
Section~\ref{sec:synth}, together with illustrative examples
showcasing the main steps of loop synthesis.
Further and most importantly, Section~\ref{sec:synth:param}
extends~\cite{HumenbergerBK20} with loop synthesis for parametrized
loops.
Section~\ref{sec:additional:examples} presents illustrative
examples from our loop synthesis experiments \revision{ and Section~\ref{sec:beyond} reports on using loop synthesis for loop equivalence and teaching formal methods.}


\section{Preliminaries}
\label{sec:prelim}

Let $\K$ be a computable field with characteristic zero. We also
assume $\K$ to be 
algebraically closed, that is, every non-constant polynomial in $\K[x]$ has at
least one root in $\K$. 
The algebraic closure $\bar{\Q}$ of the field of rational numbers $\Q$ 
is such a field; $\bar{\Q}$ is called the field of algebraic numbers. 

We denote by $\K[x_1,\dots,x_n]$ the multivariate polynomial ring with
indeterminates $x_1,\dots,x_n$. For a list $x_1,\dots,x_n$, we write $\vec{x}$
if the number of variables is known from the context or irrelevant. As $\K$ is
algebraically closed, every polynomial $p\in\K[\vec{x}]$ of degree $r$ has
exactly $r$ roots.
\ifarxiv 
Therefore, the following theorem follows immediately:
\begin{theorem}
  \label{thm:inifite-zeros}
  The zero polynomial is the only polynomial in $\K[\vec{x}]$ having infinitely
  many roots.
\end{theorem}
\fi 




\subsection{Polynomial Constraint Problem (PCP)}

A \emph{polynomial constraint} $F$ is of the form $p \bowtie 0$ where $p$ is a
polynomial in $\K[\vec{x}]$ and $\mathord{\bowtie} \in
\{<,\leq,=,\neq,\geq,>\}$. A \emph{clause} is then a disjunction ${C = F_1 \lor
\dots \lor F_m}$ of polynomial constraints. A \emph{unit clause} is a special
clause consisting of a single disjunct (i.e. $m=1$). A \emph{polynomial
constraint problem (PCP)} is then given by a set of clauses $\CC$. We say that a
variable assignment $\sigma : \{x_1,\dots,x_n\} \rightarrow \K$ satisfies a
polynomial constraint $p\bowtie0$ if $p(\sigma(x_1),\dots,\sigma(x_n)) \bowtie
0$ holds. Furthermore, $\sigma$ satisfies a clause $F_1 \lor \cdots \lor F_m$ if
for some $i$, $F_i$ is satisfied by $\sigma$. Finally, $\sigma$ satisfies a
clause set -- and is therefore a solution of the PCP -- if every clause within
the set is satisfied by $\sigma$. We write $\CC\polysin\K[\vec{x}]$ to indicate
that all polynomials in the clause set $\CC$ are contained in $\K[\vec{x}]$. For
a matrix $M$ with entries $m_1,\dots,m_s$ we define the clause set $\cstr(M)$ to
be ${\{m_1=0,\dots,m_s=0\}}$.


\subsection{Number Sequences and Recurrence Relations}
\label{sec:cfinite}

A number sequence $\seq{x(n)}$ is called \emph{C-finite} if it satisfies a
linear recurrence with constant coefficients, also known as C-finite
recurrence~\cite{KauersP11}. Let $c_0,\dots,c_{r-1} \in\K$ and $c_0\neq0$, then
\begin{equation}
  \label{eq:cfinite-form}
  x(n+r) + c_{r-1}x(n+r-1) + \cdots + c_1x(n+1) + c_0x(n) = 0
\end{equation}
is a C-finite recurrence of \emph{order} $r$. The order of a sequence is defined
by the order of the recurrence it satisfies. We refer to a recurrence
of order $r$ also as an $r$-order recurrence, for example as a first-order
recurrence when $r=1$ or a second-order recurrence when $r=2$. 
A recurrence of order $r$ and $r$ initial values define a sequence, and
different initial values lead to different sequences. For simplicity, we write
$\seq{x(n)} = 0$ for $\seq{x(n)} = \seq{0}$.

\begin{example}
  Let $a\in\K$. The constant sequence $\seq{a}$ satisfies a first-order
  recurrence equation ${x(n+1) = x(n)}$ with ${x(0)=a}$. The geometric sequence
  $\seq{a^n}$ satisfies ${x(n+1) = a x(n)}$ with ${x(0)=1}$. The sequence
  $\seq{n}$ satisfies a second-order recurrence ${x(n+2) = 2 x(n+1) - x(n)}$
  with ${x(0)=0}$ and ${x(1)=1}$.\qed
\end{example}

From the closure properties of C-finite sequences~\cite{KauersP11}, the product and
the sum of C-finite sequences are also
C-finite. Moreover, we also have the following properties:

\ifarxiv 

\begin{theorem}[\cite{KauersP11}]
  \label{thm:cfinite-closure}
  Let $\seq{u(n)}$ and $\seq{v(n)}$ be C-finite sequences of order $r$ and $s$,
  respectively. Then:
  \begin{enumerate}
    \item $\seq{u(n) + v(n)}$ is C-finite of order at most $r+s$, and
    \item $\seq{u(n)\cdot v(n)}$ is C-finite of order at most $rs$.\qed
  \end{enumerate}
\end{theorem}

\begin{theorem}[\cite{KauersP11}]
  \label{thm:cfinite-zero}
  Let $\geom_1,\dots,\geom_t\in\K$ be pairwise distinct and
  $p_1,\dots,p_t\in\K[x]$. The number sequence $\seq{p_1(n)\geom_1^n + \cdots +
  p_t(n)\geom_t^n}$ is the zero sequence if and only if the sequences
  $\seq{p_1(n)},\dots,\seq{p_t(n)}$ are zero.\qed
\end{theorem}

\fi 

\begin{theorem}[\cite{KauersP11}]
  \label{thm:poly-zero}
  Let $p = c_0 + c_1 x + \cdots + c_kx^k \in\K[x]$. Then
  ${\seq{p(n)} = 0}$ if and only if $c_0 = \cdots = c_k = 0$.\qed
\end{theorem}

\begin{theorem}[\cite{KauersP11}]
  \label{thm:finite-init-values}
  Let $\seq{u}$ be a sequence satisfying a C-finite recurrence of order $r$.
  Then, $u(n) = 0$ for all $n\in\N$ if and only if $u(n) = 0$ for
  $n\in \{0,\dots,r-1\}$.\qed
\end{theorem}

We define a \emph{system of C-finite recurrences} of \emph{order} $r$ and \emph{size} $s$
to be of the form
\begin{equation*}
  X_{n+r} + C_{r-1}X_{n+r-1} + \cdots + C_1X_{n+1} + C_0X_n = 0
\end{equation*}
where $X_n = \begin{pmatrix} x_1(n) & \cdots & x_s(n) \end{pmatrix}^\intercal$
and $C_i\in\K^{s\times s}$. Every C-finite recurrence system can be transformed
into a first-order system of recurrences by increasing the size such
that we get
\begin{equation}
  \label{eq:cfinite-rec}
  X_{n+1} = B X_{n} \qquad \text{where $B$ is invertible.}
\end{equation}
The closed form solution of a C-finite recurrence system~(\ref{eq:cfinite-rec}) is determined by the roots $\geom_1,\dots, \geom_t$ of the
characteristic polynomial of $B$, or equivalently by the eigenvalues
$\geom_1,\dots, \geom_t$ of $B$.
We recall that the
characteristic polynomial $\chi_B$ of the matrix $B$ is defined as
\begin{equation}\label{eq:cfinite:charPoly}
  \chi_B(\geom) = \det(\geom I - B),
\end{equation}
where $\det$ denotes the (matrix)
determinant and $I$ the identity matrix.
Let $m_1,\dots,m_t$ respectively denote the 
multiplicities of the roots $\geom_1,\dots, \geom_t$ of $\chi_B$. 
%
The closed form of~\eqref{eq:cfinite-rec} is then given by
\begin{equation}
  \label{eq:cfinite-cf}
  X_n = \sum_{i=1}^t \sum_{j=1}^{m_i} C_{ij} \geom_i^n n^{j-1} \qquad \text{with $C_{ij} \in \K^{s\times 1}$.}
\end{equation}
However, not every choice of the $C_{ij}$
gives rise to a solution. For obtaining a solution, we substitute the general
form~(\ref{eq:cfinite-cf}) into the original system~(\ref{eq:cfinite-rec}) and
compare coefficients. \ifarxiv The following example illustrates the procedure for
computing closed form solutions.\fi

\ifarxiv 

\begin{example}
  \label{ex:fibonacci}
  The most well-known C-finite sequence is the Fibonacci sequence satisfying a
  recurrence of order $2$ which corresponds to the following first-order
  recurrence system:
  \begin{equation}
    \label{eq:ex:cfinite1}
    \begin{pmatrix}
      f(n+1) \\
      g(n+1)
    \end{pmatrix}=
    \begin{pmatrix}
      1 & 1 \\
      1 & 0
    \end{pmatrix}
    \begin{pmatrix}
      f(n) \\
      g(n)
    \end{pmatrix}
  \end{equation}
  The eigenvalues of $B$ are given by $\geom_{1,2}=\frac{1}{2}(1 \pm
  \sqrt{5})$ with multiplicities $m_1 = m_2 = 1$. Therefore, the general
  solution for the recurrence system is of the form
  \begin{equation}
    \label{eq:ex:cfinite2}
    \begin{pmatrix}
      f(n) \\
      g(n)
    \end{pmatrix}=
    \begin{pmatrix}
      c_1 \\
      c_2
    \end{pmatrix}\geom_1^n +
    \begin{pmatrix}
      d_1 \\
      d_2
    \end{pmatrix}\geom_2^n
    .
  \end{equation}
  By substituting \eqref{eq:ex:cfinite2} into \eqref{eq:ex:cfinite1}, we get the
  following constraints over the coefficients:
  \begin{equation*}
    \begin{pmatrix}
      c_1 \\
      c_2
    \end{pmatrix}\geom_1^{n+1} +
    \begin{pmatrix}
      d_1 \\
      d_2
    \end{pmatrix}\geom_2^{n+1}
    =
    \begin{pmatrix}
      1 & 1 \\
      1 & 0
    \end{pmatrix}
    \left(
    \begin{pmatrix}
      c_1 \\
      c_2
    \end{pmatrix}\geom_1^n +
    \begin{pmatrix}
      d_1 \\
      d_2
    \end{pmatrix}\geom_2^n
    \right)
  \end{equation*}
  Bringing everything to one side yields:
  \begin{equation*}
    \begin{pmatrix}
      c_1\geom_1 - c_1 - c_2 \\
      c_2\geom_1 - c_1
    \end{pmatrix}\geom_1^{n} +
    \begin{pmatrix}
      d_1\geom_2 - d_1 - d_2\\
      d_2\geom_2 - d_1
    \end{pmatrix}\geom_2^{n}
    = 0
  \end{equation*}
  For the above equation to hold, the coefficients of the $\geom_i^n$ have to be
  $0$. That is, the following linear system determines $c_1,c_2$ and $d_1,d_2$:
  \begin{equation*}
    \begin{pmatrix}
      \geom_1 - 1 & -1 & 0 & 0 \\
      -1 & \geom_1 & 0 & 0 \\
      0 & 0 & \geom_2 - 1 & -1 \\
      0 & 0 & -1 & \geom_2
    \end{pmatrix}
    \begin{pmatrix}
      c_1 \\
      c_2 \\
      d_1 \\
      d_2
    \end{pmatrix}
    = 0
  \end{equation*}
  The solution space is generated by $(\geom_1,1,0,0)$ and $(0,0,\geom_2,1)$.
  The solution space of the C-finite recurrence system hence consists of linear
  combinations of
  \begin{equation*}
    \begin{pmatrix}
      \geom_1 \\
      1
    \end{pmatrix}\geom_1^n
    \quad\text{and}\quad
    \begin{pmatrix}
      \geom_2 \\
      1
    \end{pmatrix}\geom_2^n.
  \end{equation*}
  That is, by solving the linear system
  \begin{align*}
    \begin{pmatrix}
      f(0) \\
      g(0)
    \end{pmatrix}&=
    E
    \begin{pmatrix}
      \geom_1 \\
      1
    \end{pmatrix}\geom_1^0 +
    F
    \begin{pmatrix}
      \geom_2 \\
      1
    \end{pmatrix}\geom_2^0\\
    \begin{pmatrix}
      f(1) \\
      g(1)
    \end{pmatrix}=
    \begin{pmatrix}
      1 & 1 \\
      1 & 0
    \end{pmatrix}
    \begin{pmatrix}
      f(0) \\
      g(0)
    \end{pmatrix}&=
    E
    \begin{pmatrix}
      \geom_1 \\
      1
    \end{pmatrix}\geom_1^1 +
    F
    \begin{pmatrix}
      \geom_2 \\
      1
    \end{pmatrix}\geom_2^1
  \end{align*}
  for ${E, F\in\K^{2\times 1}}$ with $f(0)=1$ and $g(0)=0$, we get
  closed forms for~\eqref{eq:ex:cfinite1}: 
  \begin{equation*}
    f(n) = \frac{5+\sqrt{5}}{5(1+\sqrt{5})}\geom_1^{n+1} - \frac{1}{\sqrt{5}}\geom_2^{n+1}
    ~\text{and}~
    g(n) = \frac{1}{\sqrt{5}}\geom_1^{n} - \frac{1}{\sqrt{5}}\geom_2^{n}
  \end{equation*}
  Then $f(n)$ represents the Fibonacci sequence starting at $1$ and $g(n)$
  starts at $0$. Solving for $E$ and $F$ with symbolic $f(0)$ and $g(0)$ yields
  a parameterized closed form, where the entries of $E$ and $F$ are linear
  functions in the symbolic initial values.
\end{example}

\fi 


\section{Our Programming Model}\label{sec:model}

Given a polynomial relation ${p(x_1,\dots,x_s)=0}$, our loop synthesis procedure
generates a first-order C-finite/affine recurrence system~\eqref{eq:cfinite-rec} with ${X_n = \begin{pmatrix} x_1(n) & \cdots &
x_s(n) \end{pmatrix}^\intercal}$, such that ${p(x_1(n),\dots,x_s(n))=0}$ holds
for all ${n\in\N}$. It is not hard to argue that every first-order C-finite
recurrence system corresponds to a loop with simultaneous variable assignments
of the following form 
%
%
\begin{equation}
  \label{eq:loop}
  \begin{tabular}{l}
  $(x_1,\dots,x_s) \gets (a_1,\dots,a_s)$\\
  $\WHILE~true~\DO$\\
  \quad$(x_1,\dots,x_s) \gets (p_1(x_1,\dots,x_s),\dots,p_s(x_1,\dots,x_s))$\\
  $\END$
  \end{tabular}
\end{equation}
where the program variables $x_1,\dots,x_s$ are numeric, $a_1,\dots,a_s$ are
(symbolic) constants in $\K$ and $p_1,\dots,p_s$ are polynomials in
$\K[x_1,\dots,x_s]$. For a loop variable $x_i$, we denote by $x_i(n)$ the value
of $x_i$ at the $n$th loop iteration. That is, we view loop variables $x_i$ as
sequences $\seq{x_i(n)}$. 
We call a loop of the form~\eqref{eq:loop} \emph{parameterized} if at
least one of $a_1,\dots,a_s$ is symbolic, and \emph{non-parameterized}
otherwise.

\begin{remark} 
  Our synthesized loops of the form~\eqref{eq:loop} are non-deterministic, with
  loop guards being $true$. We synthesize loops such that the given
  invariant holds for an arbitrary/unbounded number of loop
  iterations -  for example, also for loop guards $n<N$ as in
  Figure~\ref{fig:Dafny}. \qed
\end{remark}


%
\begin{remark}\label{rem:petter}
  While the output of our synthesis procedure is basically an affine program, 
  note that C-finite recurrences capture a larger class of programs.
  E.g.~the program:
  \[(x,y) \gets (0,0);~\WHILE~true~\DO~(x,y) \gets (x+y^2,y+1)~\END\]
  can be modeled by a C-finite recurrence system of order $4$, which can be
  turned into an equivalent first-order system of size $6$. Thus, to
  synthesize loops inducing the sequences $\seq{x(n)}$ and
  $\seq{y(n)}$, 
  we have to consider recurrence systems of size $6$.\qed
\end{remark}
%


\ifarxiv 
\begin{example}
  \label{ex:fibonacci-loop}
  The Fibonacci recurrence system~\eqref{eq:ex:cfinite1} in Example~\ref{ex:fibonacci}
  corresponds to the following loop:
  \[(f,g) \gets (1,0);~\WHILE~true~\DO~(f,g) \gets (f+g,f)~\END \hfill\qed\]
\end{example}
\fi 


\paragraph{Algebraic relations and loop invariants.} 
Let $p$ be a polynomial in $\K[z_1,\dots,z_s]$ and let
$(x_1(n))^\infty_{n=0},\dots,(x_s(n))^\infty_{n=0}$ be number sequences. We call
$p$ an \emph{algebraic relation} for the given sequences if
${p(x_1(n),\dots,x_s(n)) = 0}$ for all ${n\in\N}$. Moreover, $p$ is an algebraic
relation for a system of recurrences if it is an algebraic relation for the
corresponding sequences. It is immediate that for every algebraic relation $p$
of a recurrence system, ${p=0}$ is a \emph{loop invariant} for the corresponding
loop of the form~\eqref{eq:loop}; that is, ${p=0}$ holds before and after every
loop iteration.


\section{Algebra-based Loop Synthesis}
\label{sec:synth}

We now present our approach for synthesizing loops satisfying a given polynomial
property (invariant), by using affine loop assignments. We transform the loop synthesis problem into
a PCP as described in 
Section~\ref{sec:synth:overview}. In Section~\ref{sec:synth:nonparam}, we
introduce the clause sets of our PCP which precisely describe the solutions for the
synthesis of loops, in particular to  non-parameterized loops. 
\ifarxiv
We extend this approach in Section~\ref{sec:synth:param} to parameterized loops.
\fi


\newcommand{\Croots}{\CC_\mathsf{roots}}
\newcommand{\Ccoeff}{\CC_\mathsf{coeff}}
\newcommand{\Cinit}{\CC_\mathsf{init}}
\newcommand{\Calg}{\CC_\mathsf{alg}}

\subsection{Setting and Overview of Our Method}
\label{sec:synth:overview}

Given a constraint ${p=0}$ with $p\in\K[x_1,\dots,x_s,y_1,\dots,y_s]$,
we aim 
to synthesize a system of C-finite recurrences such that $p$ is an algebraic
relation thereof. Intuitively, the values of loop variables $x_1,\dots,x_s$ are
described by the sequences $x_1(n),\dots,x_s(n)$ for arbitrary $n$, and
$y_1,\dots,y_s$ correspond to the initial values $x_1(0),\dots,x_s(0)$. That is,
we have a polynomial relation $p$ among loop variables $x_i$
and their initial values $y_i$, for which we synthesize a loop of the form~\eqref{eq:loop} such that ${p=0}$
is a loop invariant of a loop defined in~\eqref{eq:loop}.

\begin{remark}
  Our approach is not limited to invariants describing relationship between
  program variables from a single loop iteration. Instead, it naturally extends
  to relations among different loop iterations. For instance, by considering the
  relation in equation~\eqref{eq:fib:relation}, we synthesize a loop computing
  the Fibonacci sequence.\qed
\end{remark}

\begin{figure}
  \centering
  \tikzstyle{inner} = [rectangle, rounded corners=1mm, inner xsep=2mm, text width=1.8cm, dashed, minimum width=1.7cm, minimum height=1cm, text centered, draw=black] 
  \tikzstyle{input} = [rectangle, rounded corners=1mm, inner xsep=2mm, text width=1.8cm, minimum width=1.7cm, minimum height=1cm, text centered, draw=black] 
  \tikzstyle{arrow} = [thick,->,>=stealth]
  \begin{tikzpicture}[node distance=1.5cm]
    \node (cfsys) [inner] {Closed form system};
    \node (poly) [input, left of=cfsys, xshift=-1.8cm] {Polynomial invariant};
    \node (recsys) [inner, right of=cfsys, xshift=2.9cm] {Recurrence system};
    \node (loop) [right of=recsys, xshift=1.4cm] {Loop};
  
    \draw (poly) -- node[anchor=south] {$\Calg$} (cfsys);
    \draw [arrow] (recsys) -- (loop);
    \draw (cfsys) -- node[anchor=south] {$\Croots$, $\Ccoeff$} node[anchor=north] {$\Cinit$} (recsys);
  \end{tikzpicture}
  \caption{Overview of the PCP 
    describing loop synthesis}
  \label{fig:overview}
\end{figure}
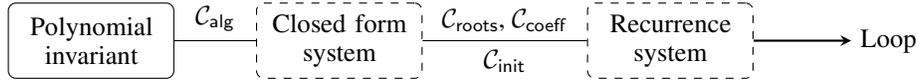

The key step in our work comes with precisely capturing the solution
space for our loop synthesis problem as a PCP.
Our PCP is divided into the clause sets $\Croots$, $\Ccoeff$, $\Cinit$
and $\Calg$, as illustrated in Figure~\ref{fig:overview} 
and explained next. Our PCP implicitly describes a first-order C-finite
recurrence system and its corresponding closed form system. The one-to-one
correspondence between these two systems is captured by the clause sets
$\Croots$, $\Ccoeff$ and $\Cinit$. Intuitively, these constraints mimic the
procedure for computing the closed form of a recurrence system
(see~\cite{KauersP11}). The clause set $\Calg$ interacts between the closed form
system and the polynomial constraint ${p=0}$, and ensures that $p$ is an
algebraic relation of the system. Furthermore, the recurrence system is
represented by the matrix $B$ and the vector $A$ of initial values where both
consist of symbolic entries. Then a solution of our PCP -- which assigns values
to those symbolic entries -- yields a desired synthesized loop.

In what follows we only consider a unit constraint $p=0$ as input to our loop
synthesis procedure. However, our approach naturally extends to conjunctions of polynomial
equality constraints.




\subsection{Synthesizing Non-Parameterized Loops}
\label{sec:synth:nonparam}

We now present our work for synthesizing loops, in particular 
non-parameterized loops of the form~\eqref{eq:loop}. That is, we aim at computing concrete initial values
for all program variables. Our implicit representation of the recurrence system
is thus of the form 
\begin{equation}
  \label{eq:nonparam:recsys}
  X_{n+1} = BX_n,  \qquad X_0 = A
\end{equation}
where $B\in\K^{s\times s}$ is invertible and $A\in\K^{s\times 1}$, both
containing symbolic entries.

As described in Section~\ref{sec:cfinite}, the closed form
of~\eqref{eq:nonparam:recsys}
is determined by the roots of the
characteristic polynomial $\chi_B(\geom)$ defined
in~\eqref{eq:cfinite:charPoly}. Using the closed form
representation~\eqref{eq:cfinite-cf}, we conclude that the eigenvalues 
of the coefficient matrix $B$ in~\eqref{eq:nonparam:recsys}
determine the closed form
of~\eqref{eq:nonparam:recsys}, implying thus the need of synthesizing the (symbolic) eigenvalues of
$B$. 
Note that $B$ may contain both symbolic and concrete
values \revision{(where concrete values may come from user-given initial values or additional user-given values for the symbolic parameters of the loop given in equation~\eqref{eq:loop}).} Let us denote the symbolic entries of $B$ by $\vec{b}$. 
Since $\K$ is algebraically
closed, we know that $B$ has $s$ (not necessarily distinct) eigenvalues.
We therefore fix a set of distinct
symbolic eigenvalues $\geom_1,\dots,\geom_t$ together with their multiplicities
$m_1,\dots,m_t$ with $m_i>0$ for $i=1,\dots,t$ such that $\sum_{i=1}^t m_i = s$. 
We call $m_1,\dots,m_t$ an \emph{integer partition} of $s$.
We next define the clause sets of our PCP. 

\paragraph{Root constraints $\Croots$.}

The clause set $\Croots$ ensures that $B$ is invertible and that
$\geom_1,\dots,\geom_t$ are distinct symbolic eigenvalues with multiplicities
$m_1,\dots,m_t$. Note that $B$ is invertible if and only if all eigenvalues
$\geom_i$ are non-zero. Furthermore, since $\K$ is algebraically closed, every
polynomial $f(z)$ can be written as the product of linear factors of the form $z
- \geom$, with $\geom\in\K$, such that $f(\geom)=0$. Let us recall that the
characteristic polynomial $\chi_B$ of the matrix $B$ is defined as
${\chi_B(\geom) = \det(\geom I - B)}$, where $\det$ denotes the (matrix)
determinant and $I$ the identity matrix. Then, the equation 
\begin{equation}\label{eq:groupZ}
  \chi_B(z) = (z - \geom_1)^{m_1} \cdots (z - \geom_t)^{m_t}
\end{equation}
holds for all $z\in\K$, where $\chi_B(z)\in\K[\vec{\geom},\vec{b},z]$.
Bringing everything to one side and \revision{regrouping terms based on the monomials in $z$}, we get 
\begin{equation*}
  q_0 + q_1z + \cdots + q_d z^d = 0, 
\end{equation*}
implying that the ${q_i\in\K[\vec{\geom},\vec{b}]}$ have to be zero. \revision{Note that $q_i$ are the coefficients of the monomials in $z$ resulting from the equation~\eqref{eq:groupZ}; as such, $q_i$ are polynomials in $\vec{\geom},\vec{b}$.}  
The clause set characterizing the eigenvalues $\geom_i$ of $B$ is then
\begin{align*}
  \Croots = \{ q_0=0,\dots,q_d=0 \}
  \cup \bigcup_{\substack{i,j=1,\dots,t\\i\neq j}} \{\geom_i \neq
  \geom_j\} \cup \bigcup_{i=1,\dots,t} \{\geom_i \neq 0\}.
\end{align*}
%

\paragraph{Coefficient constraints $\Ccoeff$.}

The fixed symbolic roots/eigenvalues $\geom_1,\dots,\geom_t$ with multiplicities
$m_1,\dots,m_t$ induce the general closed form solution
\begin{equation}
  \label{eq:generalform}
  X_n = \sum_{i=1}^t \sum_{j=1}^{m_i} C_{ij} \geom_i^n n^{j-1}
\end{equation}
where the ${C_{ij}\in\K^{s\times 1}}$ are column vectors containing symbolic
entries. As stated in Section~\ref{sec:cfinite}, not every choice of
the $C_{ij}$ gives rise to a valid solution. Instead,  $C_{ij}$ have to obey certain
conditions which are determined by substituting into the original recurrence
system of \eqref{eq:nonparam:recsys}:
\begin{align*}
  X_{n+1} 
  &= \sum_{i=1}^t \sum_{j=1}^{m_i} C_{ij} \geom_i^{n+1}(n+1)^{j-1}
  = \sum_{i=1}^t \sum_{j=1}^{m_i} \left(\sum_{k=j}^{m_i} \binom{k-1}{j-1} C_{ik} \geom_i\right) \geom_i^{n}n^{j-1} \\
  &= B \left(\sum_{i=1}^t \sum_{j=1}^{m_i} C_{ij} \geom_i^n n^{j-1}\right) = BX_n
\end{align*}
Bringing everything to one side yields ${X_{n+1} - BX_n = 0}$ and thus
\begin{equation}
  \label{eq:cstr-cf}
  \sum_{i=1}^t \sum_{j=1}^{m_i} \underbrace{\left(\left(\sum_{k=j}^{m_i} \binom{k-1}{j-1} C_{ik} \geom_i\right) - BC_{ij}\right)}_{D_{ij}} \geom_i^{n}n^{j-1} = 0.
\end{equation}
Equation~\eqref{eq:cstr-cf} holds for all $n\in\N$. By 
Theorem~\ref{thm:poly-zero} we then have $D_{ij} = 0$ for all $i,j$
and define  
%
%
\begin{equation*}
  \Ccoeff = \bigcup_{i=1}^t \bigcup_{j=1}^{m_i} \cstr(D_{ij}).
\end{equation*}
%

\paragraph{Initial values constraints $\Cinit$.}
The constraints $\Cinit$ describe properties of initial values
$x_1(0),\dots,x_s(0)$. We enforce that \eqref{eq:generalform} equals
$B^nX_0$, for $n=0,\dots,d-1$, where $d$ is the degree of the characteristic
polynomial $\chi_B$ of $B$, by 
\begin{equation*}
  \label{eq:initvalues}
  \Cinit = \cstr(M_0) \cup\cdots\cup \cstr(M_{d-1})
\end{equation*}
where $M_i = X_i - B^iX_0$, with ${X_0=A}$ as
in~\eqref{eq:nonparam:recsys} and $X_i$ 
being the right-hand side of~\eqref{eq:generalform} where $n$ is replaced by
$i$.
%

\paragraph{Algebraic relation constraints $\Calg$.}
The constraints $\Calg$ are defined to ensure that $p$ is an
algebraic relation among the $x_i(n)$. Using~\eqref{eq:generalform}, the
closed forms of the $x_i(n)$ are expressed as
\begin{equation*}
  \label{eq:cfpoly}
  x_i(n) = p_{i,1} \geom_1^n + \cdots + p_{i,t} \geom_t^n
\end{equation*}
where the $p_{i,j}$ are polynomials in $\K[n,\vec{c}]$. By substituting the
closed forms and the initial values into the polynomial $p$, we get
\begin{equation}
  \label{eq:invpoly}
  \begin{aligned}
  p' 
  = p(x_1(n),\dots,x_s(n),x_1(0),\dots,x_s(0)) 
  = q_0 + nq_1 + n^2q_2 + \cdots + n^kq_k
  \end{aligned}
\end{equation}
where the $q_i$ are of the form
\begin{equation}
  \label{eq:qform}
  w_{i,1}^nu_{i,1} + \cdots + w_{i,{\ell}}^n u_{i,{\ell}}
\end{equation}
with $u_{i,1},\dots,u_{i,{\ell}}\in\K[\vec{a},\vec{c}]$ and
$w_{i,1},\dots,w_{i,{\ell}}$ being monomials in $\K[\vec{\geom}]$. 

\begin{proposition}
  \label{prop:coeff-zero}
  Let $p$ be of the form~\eqref{eq:invpoly}. Then ${\seq{p(n)}=0}$ iff
  ${\seq{q_i(n)} = 0}$ for ${i=0,\dots,k}$. \qed
\end{proposition}

\ifarxiv 
\begin{proof}
  One direction is obvious and for the other assume ${p(n)=0}$. By rearranging
  $p$ we get $p_1(n) w_1^n + \cdots + p_\ell(n) w_\ell^n$. Let
  $\tilde{\geom}_1,\dots,\tilde{\geom}_t\in\K$ be such that $\tilde{p} = p_1(n)
  \tilde{w}_1^n + \cdots + p_\ell(n) \tilde{w}_\ell^n = 0$ with $\tilde{w}_i =
  w_i(\vec{\tilde\geom})$. Note that the $\tilde{w}_i$ are not necessarily
  distinct. However, consider $v_1,\dots,v_r$ to be the pairwise distinct elements
  of the $\tilde{w}_i$. Then we can write $\tilde{p}$ as $\sum_{i=1}^r v_i^n
  (p_{i,0} + n p_{i,1} + \cdots + n^k p_{i,k})$. By
  Theorems~\ref{thm:cfinite-zero} and~\ref{thm:poly-zero} we get that the
  $p_{i,j}$ have to be $0$. Therefore, also $v_i^n p_{i,j} = 0$ for all $i,j$.
  Then, for each $j=0,\dots,k$, we have ${v_1^np_{1,j} + \cdots + v_r^np_{1,j} = 0
  = q_j}$.
\end{proof}
\fi 

As $p$ is an algebraic relation, we have that $p'$ should be $0$
for all $n\in\N$. Proposition~\ref{prop:coeff-zero} then implies that the $q_i$
have to be $0$ for all $n\in\N$.
\begin{lemma}
  \label{lemma:cfinite}
  Let $q$ be of the form~\eqref{eq:qform}. Then ${q=0}$ for all ${n\in\N}$ if
  and only if ${q=0}$ for ${n\in\{0,\dots,\ell-1\}}$. \qed
\end{lemma}

\ifarxiv 
\begin{proof}
  The proof follows from Theorem~\ref{thm:finite-init-values} and from the fact
  that $q$ satisfies a C-finite recurrence of order $l$. To be more precise, the
  $u_{i,j}$ and $w_{i,j}^n$ satisfy a first-order C-finite recurrence: as
  $u_{i,j}$ is constant it satisfies a recurrence of the form $x(n+1)=x(n)$, and
  $w_{i,j}^n$ satisfies $x(n+1) = w_i x(n)$. Then, by
  Theorem~\ref{thm:cfinite-closure} we get that $w_{i,j}^n u_{i,j}$ is C-finite
  of order at most $1$, and $q$ is C-finite of order at most $\ell$.
\end{proof}
\fi 

Even though the $q_i$ contain exponential terms in $n$, it follows from
Lemma~\ref{lemma:cfinite} that the solutions for the $q_i$ being $0$ for all
$n\in\N$ can be described as a finite set of polynomial equality
constraints: 
Let $Q_i^j$ be the polynomial constraint ${w_{i,1}^ju_{i,1} + \cdots +
w_{i,\ell}^j u_{i,\ell} = 0}$ for $q_i$ of the form~\eqref{eq:qform}, and let ${\CC_i
= \{Q_i^0,\dots,Q_i^{\ell-1}\}}$ be the associated clause set. Then the clause set
ensuring that $p$ is indeed an algebraic relation is given by
\begin{equation*}
  \Calg = \CC_0 \cup \cdots \cup \CC_k.
\end{equation*}
%


\begin{remark}
  Observe that Theorem~\ref{thm:finite-init-values} can be applied
  to formula~\eqref{eq:invpoly} directly, as $p'$ satisfies a C-finite
  recurrence. Then by the closure properties of C-finite recurrences, 
  the upper bound on the order of the recurrence which $p'$ satisfies 
  is given by $r =
  \sum_{i=0}^k 2^i \ell$. That is, by Theorem~\ref{thm:finite-init-values}, we
  would need to consider $p'$ with $n=0,\dots,r-1$, which yields a non-linear
  system with a degree of at least $r-1$. Note that $r$ depends on
  $2^i$,  which stems from the fact that $\seq{n}$ satisfies
  a recurrence of order $2$, and $n^i$ satisfies therefore a recurrence of order
  at most $2^i$. Thankfully, Proposition~\ref{prop:coeff-zero} allows us to only consider
  the coefficients of the $n^i$ and therefore lower the size of our constraints.\qed
\end{remark}

Having defined the clause sets $\Croots$, $\Ccoeff$, $\Cinit$ and
$\Calg$, we define our PCP as the union of these four clause
sets. Note that
the matrix $B$, the vector $A$, the polynomial $p$ and the multiplicities of the
symbolic roots $\vec{m} = m_1,\dots,m_t$ uniquely define the clauses discussed
above. We define our PCP to be the clause set
$\CC_{AB}^p(\vec{m})$ as follows:
\begin{equation}
  \label{eq:ccab}
  \CC_{AB}^p(\vec{m}) = \Croots \cup \Cinit \cup \Ccoeff \cup \Calg
\end{equation}
%

Recall that $\vec{a}$ and $\vec{b}$ are the symbolic entries in the matrices $A$
and $B$ given in~\eqref{eq:nonparam:recsys}, $\vec{c}$ are the symbolic entries in the
$C_{ij}$ in the general form~\eqref{eq:generalform}, and $\vec{\geom}$ are the symbolic eigenvalues
of $B$. 
We then have ${\CC_{AB}^p(\vec{m})\polysin\K[\vec{\geom},\vec{a},\vec{b},\vec{c}]}$.

It is not difficult to see that the constraints in $\Calg$ determine
the
size of our PCP. 
As such, the degree and the number of terms
in the invariant have a direct impact on the size and the maximum degree of the
polynomials in our PCP. Which might not be obvious is that the
number of distinct symbolic roots influences the size and the maximum
degree of our PCP.
The more distinct roots are considered the higher is the number of terms in
\eqref{eq:qform}, and therefore more instances of \eqref{eq:qform}
have to be added to our PCP.

Consider ${p\in\K[x_1,\dots,x_s,y_1,\dots,y_s]}$, ${B\in\K^{s\times s}}$ and
${A\in\K^{s\times 1}}$, and let ${m_1,\dots,m_t}$ be an integer partition of
$\deg_\geom(\chi_B(\geom))$. We then get the following theorem:

\begin{theorem}
  \label{thm:nonparam}
  The mapping ${\sigma:\{\vec{\geom},\vec{a},\vec{b},\vec{c}\}\rightarrow\K}$ is
  a solution of ${\CC_{AB}^p(\vec{m})}$ if and only if
  $p(\vec{x},x_1(0),\dots,x_s(0))$ is an algebraic relation for ${X_{n+1} =
  \sigma(B) X_n}$ with ${X_0 = \sigma(A)}$, and the eigenvalues of $\sigma(B)$
  are $\sigma(\geom_1),\dots,\sigma(\geom_t)$ with multiplicities
  $m_1,\dots,m_t$.\qed
\end{theorem}


\SetKwData{nothing}{nothing}\SetKwData{sat}{sat}
\SetKwFunction{intpart}{IntPartitions}\SetKwFunction{solve}{Solve}

From Theorem~\ref{thm:nonparam}, we then get Algorithm~\ref{alg:nonparam} for
synthesizing the C-finite recurrence representation of a non-parameterized
loop of the form~\eqref{eq:loop}: $\intpart(s)$ returns the set of all integer
partitions of an integer $s$; and $\solve(\CC)$ returns whether the clause set
$\CC$ is satisfiable and a model $\sigma$ if so.
\revision{To this end, the clause set $\CC^p_{AB}$ in Algorithm~\ref{alg:nonparam} refers to our PCP problem defined in equation~\eqref{eq:ccab}. }
We further note that the growth of the
number of integer partitions is subexponential, and so is the complexity
Algorithm~\ref{alg:nonparam}. A more precise complexity analysis of
Algorithm~\ref{alg:nonparam} is  the subject of future investigations.

\begin{algorithm}[tb]
\DontPrintSemicolon
\SetKwInOut{Input}{Input}\SetKwInOut{Output}{Output}
\Input{A polynomial $p\in\K[x_1,\dots,x_s,y_1,\dots,y_s]$.}
\Output{A vector $A\in\K^{s\times 1}$ and a matrix $B\in\K^{s\times s}$ s.t.~$p$
is an algebraic relation of $X_{n+1} = BX_{n}$ and $X_0 = A$, if such $A$ and
$B$ exist.}
\BlankLine
$A \gets (a_{i}) \in \K^{s\times1}$ \tcp{symbolic vector}
$B \gets (b_{ij}) \in \K^{s\times s}$ \tcp{symbolic matrix}
\For{$m_1,\dots,m_t\in \intpart(s)$}{
  $\sat, \sigma \gets \solve(\CC^p_{AB}(m_1,\dots,m_t))$\label{alg:Solve}\;
  \lIf{$\sat$}{
    \Return{$\sigma(A), \sigma(B)$}
  }
}

\caption{Synthesis of a non-parameterized C-finite recurrence system}
\label{alg:nonparam}
\end{algorithm}

Finally, based on  Theorem~\ref{thm:nonparam} and on the property that
the number of integer
partitions of a given integer is finite, we obtain the following
result: 

\begin{theorem}
  \label{thm:sound-complete}
  Algorithm~\ref{alg:nonparam} is sound, and complete w.r.t.~recurrence systems
  of size $s$.\qed
\end{theorem}

The precise characterization of non-parameterized loops by non-parameterized
C-finite recurrence systems implies soundness and completeness of our approach
for non-parameterized loops from Theorem~\ref{thm:sound-complete}.
\revision{In fact, a
 recurrence system of size $s$ generated by Algorithm~\ref{alg:nonparam} gives
 rise to a non-parameterized loop with $s$ variables and at most $s-1$ auxiliary
 variables where the auxiliary variables capture the values of the program
 variables of previous loop iterations.}

\begin{example}
  \label{ex:nonparam}
  We showcase Algorithm~\ref{alg:nonparam} by synthesizing a loop from the loop
  invariant~${x=2y}$. That is, the polynomial is given by ${p = x-2y \in
  \K[x,y]}$, and we want to find a recurrence system of the following form 
  \begin{equation}
    \label{ex:eq:general-rec}
    \begin{pmatrix}
      x(n+1) \\
      y(n+1)
    \end{pmatrix}=
    \begin{pmatrix}
      b_{11} & b_{12} \\
      b_{21} & b_{22}
    \end{pmatrix}
    \begin{pmatrix}
      x(n) \\
      y(n)
    \end{pmatrix}
    \qquad\qquad
    \begin{pmatrix}
      x(0) \\
      y(0)
    \end{pmatrix}=
    \begin{pmatrix}
      a_1 \\
      a_2
    \end{pmatrix}
  \end{equation}
  The characteristic polynomial of $B$ is then given by
  \begin{equation*}
    \chi_B(\geom) = \geom^2 - b_{11}\geom - b_{22}\geom - b_{12}b_{21}
    + b_{11}b_{22} 
  \end{equation*}
  where its roots define the closed form system. Since we cannot determine the
  actual roots of $\chi_B(\geom)$, we have to fix a set of symbolic roots. The
  characteristic polynomial has two -- not necessarily distinct -- roots: Either
  $\chi_B(\geom)$ has two distinct roots $\geom_1,\geom_2$ with multiplicities
  ${m_1=m_2=1}$, or a single root $\geom_1$ with multiplicity ${m_1=2}$. Let us
  consider the latter case. The first clause set we define is $\Croots$ for
  ensuring that $B$ is invertible (i.e.~$\geom_1$ is nonzero), and that
  $\geom_1$ is indeed a root of the characteristic polynomial with multiplicity
  $2$. That is, $\chi_B(\geom) = (\geom - \geom_1)^2$ has to hold for all
  $\geom\in\K$, and bringing everything to one side yields 
  \begin{equation*}
    (b_{11} + b_{22} - 2\geom_1) \geom + b_{12}b_{21} - b_{11}b_{22} + \geom_1^2 = 0
  \end{equation*}
  %
  We then get the following clause set: 
  \begin{equation*}
    \Croots = \{ b_{11} + b_{22} - 2\geom_1 = 0, b_{12}b_{21} - b_{11}b_{22} + \geom_1^2 = 0, \geom_1\neq0 \}
  \end{equation*}
  As we fixed the symbolic roots, the general closed form system is
  of the form 
  \begin{equation}
    \label{ex:eq:general-cf}
    \begin{pmatrix}
      x(n) \\ y(n)
    \end{pmatrix} =
    \begin{pmatrix}
      c_1 \\ c_2
    \end{pmatrix} \geom_1^n +
    \begin{pmatrix}
      d_1 \\ d_2
    \end{pmatrix} \geom_1^n n
  \end{equation}
  By substituting into the recurrence system we get:
  \begin{equation*}
    \begin{pmatrix}
      c_1 \\ c_2
    \end{pmatrix} \geom_1^{n+1} +
    \begin{pmatrix}
      d_1 \\ d_2
    \end{pmatrix} \geom_1^{n+1} (n+1) =
    \begin{pmatrix}
      b_{11} & b_{12} \\
      b_{21} & b_{22}
    \end{pmatrix}
    \left(\begin{pmatrix}
      c_1 \\ c_2
    \end{pmatrix} \geom_1^n {+}
    \begin{pmatrix}
      d_1 \\ d_2
    \end{pmatrix} \geom_1^n n \right)
  \end{equation*}
  By further simplifications and re-ordering of terms  we then obtain:
  %
  %
  \begin{align*}
    0 = \begin{pmatrix}
      c_1\geom_1+d_1\geom_1 - b_{11}c_1-b_{12}c_2 \\ c_2\geom_1+d_2\geom_1 - b_{21}c_1-b_{22}c_2
    \end{pmatrix} \geom_1^{n}
    +
    \begin{pmatrix}
      d_1\geom_1 - b_{11}d_1-b_{12}d_2 \\ d_2\geom_1 - b_{21}d_1-b_{22}d_2
    \end{pmatrix} \geom_1^{n} n
  \end{align*}
  Since this equation has to hold for $n\in\N$ we get the following clause set:
  \begin{align*}
    \Ccoeff = \{
      &c_1\geom_1+d_1\geom_1 - b_{11}c_1-b_{12}c_2=0, 
      c_2\geom_1+d_2\geom_1 - b_{21}c_1-b_{22}c_2=0, \\
      &d_1\geom_1 - b_{11}d_1-b_{12}d_2=0, 
      d_2\geom_1 - b_{21}d_1-b_{22}d_2=0
    \}
  \end{align*}
  For defining the relationship between the closed forms and
  the initial values, we set the formula~\eqref{ex:eq:general-cf} with $n=i$ to be
  equal to the $i^\kth$ unrolling of formula~\eqref{ex:eq:general-rec} for $i=0,1$:
  \begin{equation*}
    \begin{pmatrix}
      c_1 \\ c_2
    \end{pmatrix} = 
    \begin{pmatrix}
      a_1 \\ a_2
    \end{pmatrix}
    \qquad
    \begin{pmatrix}
      c_1 \\ c_2
    \end{pmatrix} \geom_1 +
    \begin{pmatrix}
      d_1 \\ d_2
    \end{pmatrix} \geom_1 = 
    \begin{pmatrix}
      b_{11} & b_{12} \\
      b_{21} & b_{22}
    \end{pmatrix}
    \begin{pmatrix}
      a_1 \\ a_2
    \end{pmatrix}
  \end{equation*}
  The resulting constraints for defining the initial values are then given by
  %
  \begin{align*}
    \Cinit = \{ &c_1-a_1=0, c_1\geom_1 + d_1\geom_1 - b_{11}a_1 - b_{12}a_2 = 0,\\
                &c_2-a_2=0, c_2\geom_1 + d_2\geom_1 - b_{21}a_1 - b_{22}a_2 = 0 \}.
  \end{align*}
  %
  Eventually, we want to restrict the solutions such that $x-2y=0$ is an algebraic
  relation for our recurrence system. That is, by substituting the closed forms
  into the expression~${x(n)-2y(n)=0}$ we get
  %
  \begin{align*}
    0 &= x(n) - 2y(n) 
       = c_1\geom_1^n + d_1\geom_1^nn - 2(c_2\geom_1^n + d_2\geom_1^nn)\\
       &= \underbrace{\left(c_1-2c_2\right)\geom_1^n}_{q_0} + \underbrace{\left(\left( d_1-2d_2 \right)\geom_1^n\right)}_{q_1} n
  \end{align*}
  %
  where $q_0$ and $q_1$ have to be $0$ since the above equation has to hold for
  all $n\in\N$. Then, by applying Lemma~\ref{lemma:cfinite} to $q_0$ and $q_1$,
  we get the following clauses:
  \begin{equation*}
    \Calg = \{ c_1-2c_2=0, d_1-2d_2=0 \}
  \end{equation*}
 Our PCP is then the union of $\Croots$, $\Ccoeff$, $\Cinit$ and
  $\Calg$. Two possible solutions for our PCP,  and therefore of
  the synthesis problem, are given by the following loops:
  \begin{equation*}
    \begin{tabular}{l}
    $(x, y) \gets (2, 1)$\\
    $\WHILE~true~\DO$\\
    $\quad(x, y) \gets (x+2, y+1)$\\
    $\END$\\
    \end{tabular}
    \qquad\quad
    \begin{tabular}{l}
      $(x, y) \gets (2, 1)$\\
      $\WHILE~true~\DO$\\
      $\quad(x, y) \gets (2x, 2y)$\\
      $\END$\\
    \end{tabular}
  \end{equation*}
  Note that both loops above have mutually independent affine updates. Yet, the second
  one induces geometric sequences and requires handling exponentials of $2^n$.\qed
\end{example}

The completeness in Theorem~\ref{thm:sound-complete} is relative to systems of
size $s$ which is a consequence of the fact that we synthesize first-order
recurrence systems. That is, there exists a system of recurrence equations of order~${}>1$ and
size $s$ with an algebraic relation $p\in\K[x_1,\dots,x_s]$, but there exists no
first-order system of size $s$ where $p$ is an algebraic relation. \revision{Therefore, in
order to be able to synthesize a system satisfying $p$ as an algebraic relation, we have to look for recurrence systems of size~$>s$. Example~\ref{ex:synth:petter} next 
discusses an instance of such a higher-order recurrence system.
In order to gain completeness in Theorem~\ref{thm:sound-complete} 
independent of the size $s$ of the recurrence system, we would need to be able to derive an upper bound on $s$.  Obtaining such a bound from $p$ seems to be highly
non-trivial with connections to the inverse problem of difference Galois theory~\cite{Galois},  and is subject of future work.}

\revision{
\begin{example}
  \label{ex:synth:petter}
  Consider the loop given in Remark~\ref{rem:petter}. This loop has ${2y^3-3y^2+y-6x=0}$ as  a loop invariant, as inferred for example by~\cite{HumenbergerBK20}.

  Even though the loop assignment for $x$ in the  example from Remark~\ref{rem:petter} is nonlinear, the
  number sequences induced by $x$ and $y$ respectively satisfy the C-finite
  recurrence equations
  \begin{align*}
    x(n+4) - 4x(n+3) + 6x(n+2) - 4x(n+1) + x(n) = 0\\
    y(n+2) - 2y(n+1) + y(n) = 0
  \end{align*}
  with the following initial values of the loop variables $x$ and $y$: 
  \begin{alignat*}{8}
    x(0) &= &&= 0\\
    x(1) &= x(0) + y(0)^2 &&= 0 &\qquad\qquad\qquad y(0) &= &&= 0\\
    x(2) &= x(1) + y(1)^2 &&= 1 &\qquad y(1) &= y(0) + 1 &&= 1 \\
    x(3) &= x(2) + y(2)^2 &&= 5 
  \end{alignat*}
  We therefore have a C-finite recurrence system of order $4$ which is
  equivalent to the first-order system of recurrence equations
  \begin{equation*}
    \begin{pmatrix}
      u_0(n+1) \\
      u_1(n+1) \\
      u_2(n+1) \\
      u_3(n+1)
    \end{pmatrix}=
    \begin{pmatrix}
      0 & 1 & 0 & 0 \\
      0 & 0 & 1 & 0 \\
      0 & 0 & 0 & 1 \\
      1 & -4 & 6 & -4
    \end{pmatrix}
    \begin{pmatrix}
      u_0(n) \\
      u_1(n) \\
      u_2(n) \\
      u_3(n)
    \end{pmatrix}
    \qquad\qquad
    \begin{pmatrix}
      v_0(n+1) \\
      v_1(n+1)
    \end{pmatrix}=
    \begin{pmatrix}
      0 & 1 \\
      1 & -2
    \end{pmatrix}
    \begin{pmatrix}
      v_0(n) \\
      v_1(n)
    \end{pmatrix}
  \end{equation*}
  with initial values ${u_i(0)=x(i)}$ and $v{_i(0) = y(i)}$. Then $u_0$ and
  $v_0$ induce the sequences $x$ and $y$ respectively. 
  However, if we want to synthesize a loop that induces these  number sequences 
  $x$ and $y$ and has ${2y^3-3y^2+y-6x=0}$ as a loop invariant, we have to consider first-order recurrence systems of size $6$ and not only of order $4$.
\end{example}

}

\ifarxiv 
\subsection{Synthesizing Parameterized Loops}
\label{sec:synth:param}

We now extend the loop synthesis approach from Section~\ref{sec:synth:nonparam}
to an algorithmic approach synthesizing parameterized loops, that is, loops
which satisfy a loop invariant for arbitrary input values. Let us first consider
the following example motivating the synthesis problem of parameterized loops. 

\begin{example}
  \label{ex:eucliddiv}
  We are interested to synthesize a loop implementing Euclidean division over
  $x,y\in\K$. Following the problem specification of~\cite{knuth}\footnote{for
  $x,y\in\K$ we want to compute $q,r\in\K$ such that $x = yq + r$ holds}, a
  synthesized loop performing Euclidean division satisfies the polynomial
  invariant ${p = \ini{x} - \ini{y}q - r = 0}$, where $\ini{x}$ and $\ini{y}$
  denote the initial values of $x$ and $y$ before the loop. It is clear, that
  the synthesized loop should be parameterized with respect to $\ini{x}$ and
  $\ini{y}$.
  With this setting, input to our synthesis approach is the invariant ${p =
  \ini{x} - \ini{y}q - r = 0}$. A recurrence system performing Euclidean
  division and therefore satisfying the algebraic relation $\ini{x} - \ini{y}q -
  r$ is then given by $X_{n+1} = BX_n$ and $X_0 = A$ with a corresponding closed
  form system $X_n = A + Cn$ where: 
  \begin{equation*}
    X_n = \begin{pmatrix}x(n) \\ r(n) \\ q(n) \\ y(n) \\ t(n)\end{pmatrix}
    \quad
    A = \begin{pmatrix}\ini{x} \\ \ini{x} \\ 0 \\ \ini{y} \\ 1\end{pmatrix}
    \quad
    B =
    \begin{pmatrix}
      1 & 0 & 0 & 0 & 0 \\
      0 & 1 & 0 & -1 & 0 \\ 
      0 & 0 & 1 & 0 & 1 \\
      0 & 0 & 0 & 1 & 0 \\
      0 & 0 & 0 & 0 & 1
    \end{pmatrix}
    \quad
    C = \begin{pmatrix} 0 \\ -\ini{y} \\ 1 \\ 0 \\ 0 \end{pmatrix}
  \end{equation*}
  Here, the auxiliary variable $t$ plays the role of the constant $1$, and $x$
  and $y$ induce constant sequences. When compared to non-parameterized C-finite
  systems/loops, note that the coefficients in the above closed forms, as well
  as the initial values of variables, are functions in the parameters $\ini{x}$
  and $\ini{y}$.\qed
\end{example}

Example~\ref{ex:eucliddiv} illustrates that the parameterization has the effect
that we have to consider parameterized closed forms and initial values. For
non-parameterized loops we have that the coefficients in the closed forms are
constants, whereas for parameterized systems we have that the coefficients are
functions in the parameters -- the symbolic initial values of the sequences. In
fact, we have linear functions since the coefficients are obtained by solving a
linear system (see Example~\ref{ex:fibonacci}).

As already mentioned, the parameters are a subset of the symbolic initial values
of the sequences. Therefore, let ${I = \{k_1,\dots,k_r\}}$ be a subset of the
indices $\{1,\dots,s\}$. We then define $\ini{X} =
\begin{pmatrix} \ini{x}_{k_1} & \cdots & \ini{x}_{k_r} &
1\end{pmatrix}^\intercal$ where $\ini{x}_{k_1},\dots,\ini{x}_{k_r}$ denote the
parameters. Then, instead of~\eqref{eq:nonparam:recsys}, we get
\begin{equation}
  \label{eq:param:recsys}
  X_{n+1} = BX_n \qquad X_0 = A\ini{X}
\end{equation}
as the implicit representation of our recurrence system where the entries of
${A\in\K^{s\times r+1}}$ are defined as
\begin{equation*}
  a_{ij} = 
  \begin{cases}
    1 & i = k_j \\
    a_{ij}~\text{symbolic} & i \notin I \\
    0 & \text{otherwise}
  \end{cases}
\end{equation*}
and, as before, we have $B\in\K^{s\times s}$. Intuitively, the complex looking
construction of $A$ makes sure that we have $x_i(0)=\ini{x}_i$ for $i\in I$.
\begin{example}
  For the vector ${X_0 = \begin{pmatrix} x_1(0) & x_2(0) & x_3(0)
  \end{pmatrix}^\intercal}$, the set ${I=\{1,3\}}$ and therefore ${\ini{X} =
  \begin{pmatrix} \ini{x}_1 & \ini{x}_3 & 1 \end{pmatrix}^\intercal}$, we get
  the following matrix:
  \begin{equation*}
    A = \begin{pmatrix}
      1 & 0 & 0 \\
      a_{21} & a_{22} & a_{23} \\
      0 & 1 & 0
    \end{pmatrix}
  \end{equation*}
  Thus, $x_1(0)$ and $x_3(0)$ are set to $\ini{x}_1$ and $\ini{x}_3$
  respectively, and $x_2(0)$ is a linear function in $\ini{x}_1$ and
  $\ini{x}_3$.\qed
\end{example}
In addition to the change in the representation of the initial values, we also
have a change in the closed forms. That is, instead of~\eqref{eq:generalform} we
get
\begin{equation*}
  X_n = \sum_{i=1}^t \sum_{j=1}^{m_i} C_{ij}\ini{X} \geom_i^n n^{j-1}
\end{equation*}
as the general form for the closed form system with $C_{ij}\in\K^{s\times r+1}$.
Then $\Croots$, $\Cinit$, $\Ccoeff$ and $\Calg$ are defined analogously to
Section~\ref{sec:synth:nonparam}, and similar to the non-parameterized case we
define $\CC^p_{AB}(\vec{m},\vec{\ini{x}})$ as the union of those clause sets.
The polynomials in $\CC^p_{AB}(\vec{m},\vec{\ini{x}})$ are then in
$\K[\vec{\geom},\vec{a},\vec{b},\vec{c},\vec{\ini{x}}]$. Then, for each
$\vec{\geom},\vec{a},\vec{b},\vec{c} \in\K$ satisfying the clause set for all
$\vec{\ini{x}}\in\K$ gives rise to the desired parameterized loop, that is, we
have to solve an $\exists\forall$ problem. However, since all constraints
containing $\vec{\ini{x}}$ are polynomial equality constraints, we (repeatedly) apply
Theorem~\ref{thm:inifite-zeros} and get the following: 
Let ${p\in\K[\vec{\geom},\vec{a},\vec{b},\vec{c},\vec{\ini{x}}]}$ be a
polynomial such that ${p= p_1 q_1 + \dots + p_k q_k}$ with
${p_i\in\K[\vec{\ini{x}}]}$ and $q_i$ monomials in
$\K[\vec{\geom},\vec{a},\vec{b},\vec{c}]$. Then, Theorem~\ref{thm:inifite-zeros}
implies that the $q_i$ have to be $0$.


We therefore define the following operator $\decompose_{\vec{x}}(p)$ for
collecting the coefficients of all monomials in $\vec{x}$ in the polynomial $p$:
Let $p$ be of the form $q_0 + q_1x + \cdots + q_kx^k$, $P$ a clause and let
$\CC$ be a clause set, then:
\begin{align*}
  \decompose_{\vec{y},x}(p) &= 
  \begin{cases}
    \{q_0=0,\dots,q_k=0\} &\text{if $\vec{y}$ is empty} \\
    \decompose_{\vec{y}}(q_0)\cup\cdots\cup\decompose_{\vec{y}}(q_k) &\text{otherwise}
  \end{cases}\\
  \decompose_{\vec{y}}(P) &= 
  \begin{cases}
    \decompose_{\vec{y}}(p) &\text{if $P$ is a unit clause $p=0$} \\
    \{P\} &\text{otherwise}
  \end{cases}\\
  \decompose_{\vec{y}}(\CC) &= \bigcup_{P\in\CC} \decompose_{\vec{y}}(P)
\end{align*}
\revision{
\begin{example}
  Applying the operator $\decompose_x$ to the clause set
  \begin{align*}
    \CC = \{ &yzx^3+y^3x^2+(z^2-3)x=0, \\ &x-1=0 \lor y-1=0 \lor z-1=0,\\ &z\neq y \}
  \end{align*}
  yields
  \begin{align*}
    \decompose_x(\CC) = \{ &yz=0, y^3=0, z^2-3=0, \\ &x-1=0 \lor y-1=0 \lor z-1=0,\\ &z\neq y \}.
  \end{align*}
\end{example}
}

Note that the polynomials occurring in the constraint problem
$\decompose_{\vec{\ini{x}}}(\CC^p_{AB}(\vec{m},\vec{\ini{x}}))$ are  elements
of $\K[\vec{\geom},\vec{a},\vec{b},\vec{c}]$. That is,  ${\decompose_{\vec{\ini{x}}}(\CC^p_{AB}(\vec{m},\vec{\ini{x}}))
\polysin \K[\vec{\geom},\vec{a},\vec{b},\vec{c}]}$. Moreover, for
${p\in\K[x_1,\dots,x_s,y_1,\dots,y_s]}$, matrices $A$,$B$ and $\ini{X}$ as
in~\eqref{eq:param:recsys}, and an integer partition ${m_1,\dots,m_t}$ of
$\deg_\geom(\chi_B(\geom))$ we get the following theorem:

\begin{theorem}
  \label{thm:param}
  The map ${\sigma:\{\vec{\geom},\vec{a},\vec{b},\vec{c}\}\rightarrow\K}$ is
  a solution of
  ${\decompose_{\vec{\ini{x}}}(\CC^p_{AB}(\vec{m},\vec{\ini{x}}))}$
  if and only if $p(\vec{x},x_1(0),\dots,x_s(0))$ is an algebraic relation for
  $X_{n+1} = \sigma(B) X_n$ with $X_0 = \sigma(A)\ini{X}$, and
  $\sigma(\geom_1),\dots,\sigma(\geom_t)$ are the eigenvalues of $\sigma(B)$
  with multiplicities $m_1,\dots,m_t$.\qed
\end{theorem}

Theorem~\ref{thm:param} gives rise to an algorithm analogous to
Algorithm~\ref{alg:nonparam}. Furthermore, we get an analogous soundness and
completeness result as in Theorem~\ref{thm:sound-complete} which implies
soundness and completeness for parameterized loops.

\begin{example}
  We illustrate the construction of the constraint problem for
  Example~\ref{ex:eucliddiv}. For reasons of brevity, we consider a simplified
  system where the variables $r$ and $x$ are merged. The new invariant is then
  $\ini{r} = \ini{y}q + r$ and the parameters are given by $\ini{r}$ and
  $\ini{y}$. That is, we consider a recurrence system of size $4$ with sequences
  $y$, $q$ and $r$, and $t$ for the constant $1$.  As a consequence we have that
  the characteristic polynomial $B$ is of degree $4$, and we fix the symbolic
  root $\geom_1$ with multiplicity $4$. For simplicity, we only show how to
  construct the clause set $\Calg$.
  
  With the symbolic roots fixed, we get the following template for the system of
  closed form solutions: Let 
  \begin{equation*}
    X_n = \begin{pmatrix}r(n) & q(n) & y(n) &
    t(n)\end{pmatrix}^\intercal\qquad\text{and}\qquad V = \begin{pmatrix} \ini{r} & \ini{y} & 1
    \end{pmatrix}^\intercal,
  \end{equation*}
  and let ${C,D,E,F\in\K^{4\times 3}}$ be symbolic matrices. Then the closed
  form is given by
  \begin{align*}
    X_n &= \left(CV + DVn + EVn^2 + FVn^3\right) \geom_1^n
  \end{align*}
  and for the initial values we get
  \begin{align*}
    X_0 &=
    \begin{pmatrix}
      1 & 0 & 0 \\ a_{21} & a_{22} & a_{23} \\ 0 & 1 & 0 \\ a_{41} & a_{42} & a_{43}
    \end{pmatrix}V.
  \end{align*}
  By substituting the closed forms into the invariant ${r(0) - y(0) q(n) - r(n)
  = 0}$ and rearranging we get:
  \begin{align*}
    0 = \ini{r} &- \left(c_{21}\ini{r}\ini{y} - c_{22}\ini{y}^2 - c_{23}\ini{y} - c_{11}\ini{r} - c_{12}\ini{y} - c_{13}\right) \geom_1^n \\
                &- \left(d_{21}\ini{r}\ini{y} + d_{22}\ini{y}^2 + d_{23}\ini{y} - d_{11}\ini{r} - d_{12}\ini{y} - d_{13}\right) \geom_1^n n \\
                &- \left(e_{21}\ini{r}\ini{y} + e_{22}\ini{y}^2 + e_{23}\ini{y} - e_{11}\ini{r} - e_{12}\ini{y} - e_{13}\right) \geom_1^n n^2 \\
                &- \left(f_{21}\ini{r}\ini{y} + f_{22}\ini{y}^2 + f_{23}\ini{y} - f_{11}\ini{r} - f_{12}\ini{y} - f_{13}\right) \geom_1^n n^3
  \end{align*}
  Since the above equation should hold for all $n\in\N$ we get: 
  \begin{align*}
    \left(\ini{r}\right) 1^n - \left(c_{21}\ini{r}\ini{y} - c_{22}\ini{y}^2 - c_{23}\ini{y} - c_{11}\ini{r} - c_{12}\ini{y} - c_{13}\right) \geom_1^n &= 0 \\
    \left(d_{21}\ini{r}\ini{y} + d_{22}\ini{y}^2 + d_{23}\ini{y} - d_{11}\ini{r} - d_{12}\ini{y} - d_{13}\right) \geom_1^n &= 0 \\
    \left(e_{21}\ini{r}\ini{y} + e_{22}\ini{y}^2 + e_{23}\ini{y} - e_{11}\ini{r} - e_{12}\ini{y} - e_{13}\right) \geom_1^n &= 0\\
    \left(f_{21}\ini{r}\ini{y} + f_{22}\ini{y}^2 + f_{23}\ini{y} - f_{11}\ini{r} - f_{12}\ini{y} - f_{13}\right) \geom_1^n &= 0
  \end{align*}
  Then, by applying Lemma~\ref{lemma:cfinite}, we get:
  \begin{align*}
    \ini{r} - \left(c_{21}\ini{r}\ini{y} - c_{22}\ini{y}^2 - c_{23}\ini{y} - c_{11}\ini{r} - c_{12}\ini{y} - c_{13}\right) &= 0 \\
    \ini{r} - \left(c_{21}\ini{r}\ini{y} - c_{22}\ini{y}^2 - c_{23}\ini{y} - c_{11}\ini{r} - c_{12}\ini{y} - c_{13}\right) \geom_1 &= 0 \\
    d_{21}\ini{r}\ini{y} + d_{22}\ini{y}^2 + d_{23}\ini{y} - d_{11}\ini{r} - d_{12}\ini{y} - d_{13} &= 0\\
    e_{21}\ini{r}\ini{y} + e_{22}\ini{y}^2 + e_{23}\ini{y} - e_{11}\ini{r} - e_{12}\ini{y} - e_{13} &= 0\\
    f_{21}\ini{r}\ini{y} + f_{22}\ini{y}^2 + f_{23}\ini{y} - f_{11}\ini{r} - f_{12}\ini{y} - f_{13} &= 0
  \end{align*}
  Finally, by applying the operator $\decompose_{\ini{y},\ini{r}}$, we get the
  following constraints for $\Calg$:
  \begin{alignat*}{5}
    c_{21} &={ }& 1-c_{11} &={ }& c_{22} &={ }& c_{23} + c_{12} &={ }& c_{13} &= 0 \\
    \geom_1c_{21} &={ }& 1-\geom_1c_{11} &={ }& \geom_1c_{22} &={ }& \geom_1\left(c_{23} + c_{12}\right) &={ }& \geom_1c_{13} &= 0 \\
    d_{21} &={ }& d_{11}   &={ }& d_{22} &={ }& d_{23} + d_{12} &={ }& d_{13} &= 0 \\
    e_{21} &={ }& e_{11}   &={ }& e_{22} &={ }& e_{23} + e_{12} &={ }& e_{13} &= 0 \\
    f_{21} &={ }& f_{11}   &={ }& f_{22} &={ }& f_{23} + f_{12} &={ }& f_{13} &= 0
  \end{alignat*}\qed
\end{example}
\fi 

\section{Automating Algebra-based Loop Synthesis}\label{sec:automate}

For automating Algorithm~\ref{alg:nonparam} for loop synthesis, the challenging
task is to find solutions for our \abbrv{PCP}s describing large systems of
polynomial constraints with many variables and high polynomial degrees
(see the \abbrv{PCP} problem given in~\eqref{eq:ccab}).
We propose the following (partial)
solutions for optimizing and exploring the \abbrv{PCP} solution
space. 
 \revision{For handling recurrence templates of large size, we utilize the
flexibility of our approach by trying to find loops of a specific shape first,
and then generalize if needed. Moreover, for dealing with \abbrv{PCP}s containing
polynomials of high degree, we leverage properties of the constraints generated
by our procedure. We will see in our experimental evaluation in
Section~\ref{sec:experiments} that the former is useful in loop synthesis
and the latter is necessary for the synthesis of number sequences in a
mathematical setting.}



\paragraph{Handling large recurrence templates.}

It is obvious that the higher the number of program variables in the loop to be
synthesized is, the higher is the number of variables in the \abbrv{PCP} of
Algorithm~\ref{alg:nonparam}. To face this increase of complexity we implemented
an iterative search for \abbrv{PCP} solutions in the sense that we preset
certain values of the coefficient matrix $B$ in~\eqref{eq:nonparam:recsys}. In
particular, we start by looking for \abbrv{PCP} solutions where the coefficient
matrix $B$ is unit upper triangular. If no such solution is found, we consider
$B$ to be an upper triangular matrix and further to be full symbolic matrix
without preset values. This way we first construct simpler \abbrv{PCP}s (in
terms of the number of variables) and generalize step by step, if needed. This
iterative approach can also be used to the search for only integer \abbrv{PCP}
solutions by imposing/presetting $B$ to contain only integer-valued. 

Synthesizing a (unit) upper triangular coefficient matrix $B$ yields a loop where
its loop variables are not mutually dependent on each other. We note that such a
pattern is a very common programming paradigm -- all benchmarks from
Table~\ref{tab:experiments} satisfy such a pattern. Yet, as a consequence of
restricting the shape of $B$, the order of the variables in the recurrence
system matters. That is, we have to consider all possible variable permutations
for ensuring completeness w.r.t.~(unit) upper triangular matrices.

\paragraph{Handling large polynomial degrees.}
The main source of polynomials with high degrees in the \abbrv{PCP} of
Algorithm~\ref{alg:nonparam}
stems from the clause set $\Calg$, i.e.~constraints
of the form~\eqref{eq:qform} for $n\in\{0,\dots,\ell-1\}$.
For any \abbrv{PCP}  solution $\sigma$ in line~\ref{alg:Solve} of
Algorithm~\ref{alg:nonparam}, 
we have
${\sigma(w_{1})^n\sigma(u_{1}) + \cdots
+ \sigma(w_{{\ell}})^n \sigma(u_{{\ell}})=0}$ for
${n\in\{0,\dots,\ell-1\}}$, which yields the following system of
linear equations:
\begin{equation}
    \label{eq:vandermonde}
    W\vec{u}=
    \begin{bmatrix}
        1 & 1 & 1 & \cdots & 1 \\
        \sigma(w_1) & \sigma(w_2) & \sigma(w_3) & \cdots & \sigma(w_\ell) \\
        \sigma(w_1)^2 & \sigma(w_2)^2 & \sigma(w_3)^2 & \cdots & \sigma(w_\ell)^2 \\
        \vdots & \vdots & \vdots & \ddots & \vdots \\
        \sigma(w_1)^{\ell-1} & \sigma(w_2)^{\ell-1} & \sigma(w_3)^{\ell-1} & \cdots & \sigma(w_\ell)^{\ell-1} \\
    \end{bmatrix}
    \begin{bmatrix}
        \sigma(u_1) \\ \sigma(u_2) \\ \sigma(u_3) \\ \vdots \\ \sigma(u_\ell)
    \end{bmatrix}=0
\end{equation}
where ${W\in\K^{\ell\times\ell}}$ is a Vandermonde matrix and
${\vec{u}\in\K^\ell}$. Suppose our assignment $\sigma$ in line~\ref{alg:Solve}
of Algorithm~\ref{alg:nonparam} is such that $\sigma(w_i)\neq\sigma(w_j)$ for
$i\neq j$; if this is not the case we can always create a smaller system of the
form~\eqref{eq:vandermonde} by collecting terms. As $\sigma(w_i)\neq\sigma(w_j)$
for $i\neq j$, we derive that $W$ is invertible. Then, it follows by Cramer's
rule, that $\sigma(u_i)=0$ for all $i\in\{1,\dots,\ell\}$. Based on this
observation,  we propose Algorithm~\ref{alg:automate} for solving constraints of the
form~\eqref{eq:qform}. For simplicity, we only present the case where we have a
single constraint of the form~\eqref{eq:qform}; Algorithm~\ref{alg:automate} however naturally
extends to multiple such constraints.

\setlength{\algomargin}{0pt}
\begin{algorithm}[tb]
  \caption{Solving C-finite constraints}
\label{alg:automate}
\LinesNotNumbered

\SetKwInOut{Input}{Input}\SetKwInOut{Output}{Output}
\Input{An arbitrary satisfiable \abbrv{PCP} $\mathcal{P}$ and a constraint $C$ of the form~\eqref{eq:qform}.}
\Output{A model $\sigma$ for the polynomial constraint problem $\mathcal{P}\cup\{C\}$.}
\BlankLine
\centering
\begin{minipage}{\textwidth}
  \setlength{\leftmargin}{1pt}
\begin{enumerate}
    \item Call \textsf{MaxSAT} to compute maximum satisfiability with soft
    constraints ${u_1=0,\dots,u_\ell=0}$ and hard constraints from
    $\mathcal{P}$, and let $\sigma$ be the resulting assignment. If all soft
    constraints are satisfied, then \textbf{return} $\sigma$. Otherwise, let
    $\mathcal{I}$ be the partition such that for every set of
    indices~${I\in\mathcal{I}}$ we have ${\sigma(w_i)=\sigma(w_j)}$ for ${i,j\in
    I}$.
    \item Construct a constraint problem $\mathcal{Q}$ as follows:
    \begin{enumerate}
        \item For each $I\in\mathcal{I}$, add constraints $w_i=w_j$ for $i,j\in I$.
        \item For each distinct pair ${I,J \in \mathcal{I}}$, add a constraint
        ${w_i\neq w_j}$ for some ${i\in I}$ and ${j\in J}$.
        \item For each ${I\in\mathcal{I}}$, add a constraint ${\sum_{i\in I}u_i = 0}$.
    \end{enumerate}
    \item If $\mathcal{P}\cup\mathcal{Q}$ is satisfiable with model $\sigma$,
      then \textbf{return} $\sigma$. Otherwise, learn a new partition
      $\mathcal{I}$ and go to Step 2. 
\end{enumerate}
\end{minipage}
\BlankLine
\end{algorithm}

Intuitively, Step~1 of Algorithm~\ref{alg:automate}
finds a model $\sigma$ such that each $u_i$ becomes zero,  which makes the values of
the $w_i$ irrelevant. To this end, we compute maximum satisfiability
of our constraints $C$ using the \textsf{MaxSAT} approach
of~\cite{MaxSat14}. If this is not possible,  we continue with a partition
${\mathcal{I} = \{I_1,\dots,I_\ell\}}$ of the set of indices $\{1,\dots,\ell\}$.
Then $\mathcal{I}$ induces a system of linear equations of the
form~\eqref{eq:vandermonde} of size $\ell$ which is specified in Step~2 of the
algorithm. If the \abbrv{PCP} is satisfiable (Step~3 of Algorithm~\ref{alg:automate}), then we have found an
assignment $\sigma$ which satisfies $\mathcal{P}$ and the system of the
form~\eqref{eq:vandermonde} for $C$. If the given \abbrv{PCP} is unsatisfiable,
then we learn a new partition by making use of the unsatisfiable core and go
back to Step~2 of Algorithm~\ref{alg:automate}.

\section{Implementation and Experiments}\label{sec:experiments}

\subsection{\absynth{}:  Algebra-Based Loop Synthesis}\label{sec:impl}
Our approach to algebra-based loop synthesis is implemented in the tool
\absynth{}, which consists of about 1800 lines of Julia code and is available at
{\absynthurl}. Inputs to \absynth{} are conjunctions of polynomial equality
constraints, representing a loop invariant. As a default result, \absynth{}
derives a program that is partially correct with respect to the given invariant
(see e.g. Table~\ref{tab:experiments}). In addition, \absynth{} can also be used to
derive number sequences for which the given invariant is an algebraic relation
(Table~\ref{tab:rec-experiments}). 

As described in Section~\ref{sec:synth}, 
loop synthesis in \absynth{} is reduced to solving PCPs. These PCPs
are currently 
expressed in the quantifier-free fragment of non-linear real arithmetic
(\texttt{QF\_NRA}). We used \absynth{} in conjunction with the SMT solvers \tool{Yices}~\cite{Yices} (version 2.6.1)
and \tool{Z3}~\cite{Z3} (version 4.8.6) for solving the PCPs and
therefore synthesizing loops. 
For instance, 
Figures~\ref{fig:Dafny:b}-\ref{fig:Dafny:c} and 
Example~\ref{ex:nonparam} are synthesized automatically using \absynth{}.

As PCPs in \absynth{} are restricted to \texttt{QF\_NRA}, the
implementation of Algorithm~\ref{alg:nonparam}  within 
\absynth{} 
does not yet find solutions containing
non-real algebraic numbers. In our
loop synthesis experiments we did not encounter instances where non-real
algebraic numbers are necessary. The synthesis of recurrences, however, often
requires reasoning about non-real algebraic numbers such as the so-called Perrin
numbers $p(n)$ defined via $p(n+3)=p(n+1)+p(n)$ and satisfying the relation
$p(n)^3-3p(n)p(2n)+2p(3n) = 6$. Going beyond
the \texttt{QF\_NRA} fragment, as well as considering finite domains
(bitvectors/bounded integers) within \absynth{} is a next step to
investigate. 
%

\newcommand{\full}{\textsc{fu}}
\newcommand{\upper}{\textsc{up}}
\newcommand{\uni}{\textsc{un}}

\newcommand{\TO}{-}
\newcommand{\UN}{}

\begin{table}
  \setlength{\arraycolsep}{3pt}
  \setlength{\tabcolsep}{2.4pt}
  \fontsize{8}{8}\selectfont
  \centering
  \def\arraystretch{.95}
  \begin{tabular}{@{}lrrrrrrrrrrrrrrrrrrrr@{}}
    \toprule
    \multirow{2}{*}{Instance} & \multirow{2}{*}{{\textsc{s}}} & \multirow{2}{*}{\textsc{i}} & \multirow{2}{*}{\textsc{d}} & \multirow{2}{*}{\textsc{c}} & \phantom{x} & \multicolumn{3}{c}{\tool{Yices}} & \phantom{x} & \multicolumn{3}{c}{\tool{Z3}} & \phantom{x} & \multicolumn{3}{c}{\tool{Z3*}} & \phantom{x} & \multicolumn{3}{c}{\tool{Z3* + Alg2}} \\ 
    \cmidrule{7-9} \cmidrule{11-13} \cmidrule{15-17} \cmidrule{19-21}
    &&&&&& \uni & \upper & \full && \uni & \upper & \full && \uni & \upper & \full && \uni & \upper & \full \\
    \midrule
    \texttt{add1}*      & 5 & 1 &  5 & 173 && $932$ \UN & $921$ \UN  & \TO        && $117$ \UN & \TO       & \TO       && $22$ \UN & $726$ \UN & \TO       &&  $7$ \UN & $2416$ \UN &  \TO       \\
    \texttt{add2}*      & 5 & 1 &  5 & 173 && $959$ \UN & $861$ \UN  & \TO        && $115$ \UN & \TO       & \TO       && $22$ \UN & $109$ \UN & \TO       &&  $7$ \UN & $2323$ \UN &  \TO       \\
    \texttt{cubes}      & 5 & 3 &  6 &  94 && \TO       & \TO        & \TO        && $116$ \UN & $114$ \UN & \TO       && $18$ \UN & $496$ \UN & $575$ \UN && $87$ \UN &    \TO     &  \TO       \\
    \texttt{double1}    & 3 & 1 &  4 &  29 && $114$ \UN & $112$ \UN  & $3882$ \UN && $113$ \UN & $111$ \UN & $113$ \UN && $13$ \UN &  $21$ \UN &  $63$ \UN &&  $3$ \UN &    $5$ \UN &  $120$ \UN \\
    \texttt{double2}    & 3 & 1 &  3 &  24 && $110$ \UN & $106$ \UN  & $1665$ \UN && $115$ \UN & $106$ \UN & $115$ \UN && $13$ \UN &  $18$ \UN &  $40$ \UN &&  $2$ \UN &    $5$ \UN &   $21$ \UN \\
    \texttt{eucliddiv}* & 5 & 1 &  5 & 185 && $213$ \UN & $537$ \UN  & \TO        && $114$ \UN & $115$ \UN & \TO       && $19$ \UN &  $73$ \UN & \TO       && $10$ \UN & $2554$ \UN &  \TO       \\
    \texttt{intcbrt}*   & 5 & 2 & 12 & 262 && \TO       & \TO        & \TO        && $117$ \UN & $116$ \UN & \TO       && $22$ \UN &  $83$ \UN & $469$ \UN && $89$ \UN &    \TO     &  \TO       \\
    \texttt{intsqrt1}   & 4 & 2 &  6 &  53 && \TO       & \TO        & \TO        && $113$ \UN & $108$ \UN & $114$ \UN && $15$ \UN &  $19$ \UN & \TO       && $35$ \UN &   $81$ \UN &  \TO       \\
    \texttt{intsqrt2}*  & 4 & 1 &  6 & 104 && $105$ \UN & $1164$ \UN & \TO        && $113$ \UN & $111$ \UN & $115$ \UN && $15$ \UN &  $27$ \UN &  $37$ \UN &&  $3$ \UN &    $9$ \UN &  \TO       \\
    \texttt{petter1}    & 3 & 1 &  4 &  29 && $112$ \UN & $116$ \UN  & \TO        && $114$ \UN & $113$ \UN & $113$ \UN && $15$ \UN &  $18$ \UN &  $32$ \UN && $15$ \UN &   $32$ \UN & $3629$ \UN \\
    \texttt{square}     & 3 & 1 &  4 &  29 && $112$ \UN & $112$ \UN  & \TO        && $112$ \UN & $114$ \UN & $117$ \UN && $13$ \UN &  $17$ \UN &  $26$ \UN && $10$ \UN &   $29$ \UN &  $592$ \UN \\
    \texttt{dblsquare}  & 3 & 1 &  4 &  30 && $109$ \UN & $105$ \UN  & \TO        && $105$ \UN & $105$ \UN & $110$ \UN && $12$ \UN &  $17$ \UN &  $26$ \UN && $14$ \UN &   $31$ \UN &  \TO       \\
    \texttt{sum1}       & 4 & 2 &  6 &  53 && $617$ \UN & \TO        & \TO        && $108$ \UN & $112$ \UN & $113$ \UN && $17$ \UN &  $24$ \UN &  $99$ \UN && $39$ \UN &  $250$ \UN &  \TO       \\
    \texttt{sum2}       & 5 & 3 &  6 &  82 && \TO       & \TO        & \TO        && $220$ \UN & $112$ \UN & \TO       && $20$ \UN & $516$ \UN & \TO       && $60$ \UN &    \TO     &  \TO       \\
    \bottomrule
  \end{tabular}
  \vskip1ex
  %
  %
  \def\arraystretch{1}
  \begin{tabular}{llcll}
    \textsc{s} & size of the recurrence system & \phantom{xx} &
      {}* & parameterized system \\
    \textsc{i} & number of polynomial invariants & &
      - & timeout (60 seconds) \\
    \textsc{d} & maximum monomial degree of constraints\\
    \textsc{c} & number of constraints\\
  \end{tabular}
  \vspace{.5em}
  \caption{\absynth{} results for loop synthesis (results in milliseconds), by reverse engineering examples from the state-of-the-art in polynomial invariant generation}
  \label{tab:experiments}
\end{table}

\subsection{Experimental Results with Synthesizing Loops and Recurrences }\label{sec:exp:summaries}

Our experiments in this paper  were performed on a machine with a 2.9 GHz Intel Core i5 and 16
GB LPDDR3 RAM, and for each benchmark  a timeout of 60 seconds was set.

Tables~\ref{tab:experiments}-\ref{tab:rec-experiments}
summarize our initial experiments using academic benchmarks from the invariant generation literature~\cite{HumenbergerBK20} and recurrence solving~\cite{KauersP11}. The  columns \tool{Yices} and \tool{Z3} correspond
to the results where the respective solver is called as an external program with
and SMTLIB 2.0 file as input; \revision{as such, \tool{Yices} and \tool{Z3} are used as black-box solvers, with additional external calls on the PCP problems generated by \absynth{} as SMTLIB constraints. To reduced the overhead with external calls on solving PCP constraints, as well as to implement our PCP optimizations measures from Section~\ref{sec:automate} as guiding technologies for SMT solving over PCPs, we extended \absynth{} with SMT solving approaches tailored towards PCP reasoning. To this end, we implemented the optimizations from Section~\ref{sec:automate} as part of \absynth{}, by extending and integrating \tool{Z3} within \absynth{}.  Column \tool{Z3*} refers to the use of \absynth{} with \tool{Z3} integrated, by directly 
integrating \absynth{} with the (C\texttt{++} API) interface of 
\tool{Z3}.} Finally, column \tool{Z3* + Alg2} depicts the results
for Algorithm~\ref{alg:automate} with \tool{Z3*} as backend solver.
 The
results in Tables~\ref{tab:experiments}-\ref{tab:rec-experiments} are given in milliseconds, and only include the time needed for solving
the constraint problem as the time needed for constructing the constraints is
neglectable.

\paragraph{Loop Synthesis.} Our first benchmark set for loop synthesis
consists of invariants for loops from the
invariant generation literature~\cite{HumenbergerBK20}, and is reported in Table~\ref{tab:experiments}. Note that the benchmarks \texttt{cubes} and
\texttt{double2} in Table~\ref{tab:experiments} are those from
Figure~\ref{fig:Dafny} and Example~\ref{ex:nonparam}, respectively. A further
presentation of a selected set of our benchmarks from Table~\ref{tab:experiments}  is given in
Section~\ref{sec:additional:examples}, using the \absynth{} input language.

The columns \uni{} and \upper{} in Table~\ref{tab:experiments} show the results
where the coefficient matrix $B$ is restricted to be unit upper triangular and
upper triangular respectively. \full{} indicates that no restriction on $B$ was
set. 
Note that the running time of Algorithm~\ref{alg:nonparam} heavily depends on
the order of which the integer partitions and the variable permutations are
traversed. Therefore, in order to get comparable results, we fixed the integer
partition and the variable permutation. That is, for each instance, we enforced
that $B$ in formula~\eqref{eq:nonparam:recsys} has just a single eigenvalue, and we fixed a variable ordering where we
know that there exists a solution with an unitriangular matrix $B$. Hence, there
exists at least one solution which all cases -- \uni{}, \upper{} and \full{} --
have in common. Furthermore, for each instance we added constraints for avoiding
trivial solutions, i.e.~loops inducing constant sequences, and used
Algorithm~\ref{alg:automate} to further reduce our search space. \revision{Table~\ref{tab:experiments} shows that with these considerations on top of our optimizations from Section~\ref{sec:automate}, \tool{Z3*} outperforms \tool{Yices}. That is, 
our tailored use of \tool{Z3} directly integrated within \absynth{} brings the best performance in loop synthesis. A similar experimental conclusion can be drawn also from our experiments in using \absynth{} for synthesizing number sequences, as detailed next. }

\begin{table}
  \centering
  \begin{tabular}{@{}lrrrrrrr@{}}
    \toprule
    Instance & \textsc{o} &\phantom{x} & \tool{Yices} & \tool{Z3} & \tool{Z3*} & \tool{Z3* + Alg2} \\
    \midrule
    \texttt{fibonacci1}  & 2 && \TO & \TO & \TO & $324$ \\
    \texttt{fibonacci2}  & 2 && \TO & \TO & \TO & $22$  \\
    \texttt{example28}   & 2 && \TO & \TO & \TO & $41$  \\
    \texttt{ex1}   & 2 && \TO & \TO & \TO & $27$  \\
    \texttt{ex2}   & 2 && \TO & \TO & \TO & $20$ \\
    \texttt{ex3}   & 2 && \TO & \TO & \TO & $451$ \\
    \bottomrule
  \end{tabular}
  \vskip1ex
  %
  %
  \def\arraystretch{1}
  \begin{tabular}{llcll}
    \textsc{o} & order of recurrence & \phantom{xx} &
      \TO & timeout (60 seconds) \\
  \end{tabular}
  \vspace{.5em}
  \caption{\absynth{} results for recurrence synthesis (results in milliseconds)}
  \label{tab:rec-experiments}
\end{table}

\paragraph{Recurrence Synthesis.} In addition to loop synthesis, we also
conducted experiments with respect to synthesizing recurrence equations
(Table~\ref{tab:rec-experiments}). We took algebraic relations
from~\cite{KauersZ08} 
and synthesized recurrence
equations satisfying the given relations. None of the instances could be solved
by \tool{Yices} or \tool{Z3}, but only by \tool{Z3* + Alg2}. \revision{Based on our results from Tables~\ref{tab:experiments}-\ref{tab:rec-experiments}, we conclude that using \tool{Z3} as an integrated backend solver of \absynth{}, that is \tool{Z3*} in \absynth{}, is the right approach to take for further applications of synthesizing loops with non-linear polynomial invariants. We discuss further use cases of loop synthesis based on
\absynth{} with \tool{Z3*} in Section~\ref{sec:beyond}.
}

In contrast
to loop synthesis, we note that the synthesis of recurrence equations often requires
reasoning about non-real algebraic numbers which does not fall into the fragment
of non-linear real arithmetic. \revision{Hence, for synthesizing recurrence equations we
plan to further extend \absynth{} with a dedicate solver to reason about the whole set of
algebraic numbers; such reasoning is not yet supported by \tool{Z3*}.} 

\subsection{Examples of Synthesized Loops by \absynth{}}
\label{sec:additional:examples}
In Figure~\ref{fig:synth:experiments}, we show a few illustrative examples used in our experiments from
Table~\ref{tab:experiments}, using the \absynth{} input language. As mentioned in
Section~\ref{sec:exp:summaries}, we considered 
%
loops annotated with their invariants from the invariant generation
literature~\cite{Rodriguez-CarbonellK07,HumenbergerJK17}. 
For each
example in Figures~\ref{fig:experiments:a}-\ref{fig:experiments:e}, we first
list such a loop ({\tt Original loop}). 
We then give the first loop synthesized by our
work ({\tt \absynth{} loop})  in
combination with \tool{Yices} and \tool{Z3}, respectively\footnote{In
  Figure~\ref{fig:experiments:e}, \absynth{} with \tool{Yices} failed
  to synthesize a loop, as indicated in Table~\ref{tab:experiments}.}. That is, the
synthesized loops of
Figures~\ref{fig:experiments:a}-\ref{fig:experiments:e} are generated
by \absynth{} from the loop invariants
of~\cite{Rodriguez-CarbonellK07,HumenbergerJK17}. 
Observe that in most cases \absynth{} work was able to derive the 
loop as~\cite{Rodriguez-CarbonellK07,HumenbergerJK17}.

\newcommand{\firstcol}{0.27\textwidth}
\newcommand{\secondcol}{0.30\textwidth}
\newcommand{\thirdcol}{0.41\textwidth}
\newcommand{\captionspace}{\vspace*{-1em}}
\newcommand{\figspace}{\vspace*{.5em}}

\begin{figure}
  \captionsetup[sub]{justification=raggedright,singlelinecheck=false,margin=4em,skip=0pt}
  \begin{subfigure}{\textwidth}
\begin{center}
\begin{minipage}[t]{\firstcol}
\begin{lstlisting}[escapechar=\%]
# \textbf{eucliddiv}
# Original loop

r, q, y = x0, 0, y0
while true
  r = r - y
  q = q + 1
end
%~%
\end{lstlisting}
\end{minipage}
\begin{minipage}[t]{\secondcol}
\begin{lstlisting}[escapechar=\%]
# \textbf{eucliddiv}
# \absynth{} loop
# Solver: Yices
r, q, y = x0, 0, y0
while true
  r = r - q - y
  q = q + 1
  y = y - 1
end
\end{lstlisting}
\end{minipage}
\begin{minipage}[t]{\thirdcol}
\begin{lstlisting}[escapechar=\%]
# \textbf{eucliddiv}
# \absynth{} loop
# Solver: Z3
r, q, y = x0 - 1/2 y0, 1/2, y0
while true
  r = r - q - 1/2 y + 1/2
  q = q + 1/2
  y = y - 1
end
\end{lstlisting}
\end{minipage}
\end{center}

\captionspace
\caption{Example \texttt{eucliddiv} with loop invariant \texttt{x0 == y0*q+r}}
\label{fig:experiments:a}
\figspace
\end{subfigure}
\begin{subfigure}{\textwidth}
\begin{center}
\begin{minipage}[t]{\firstcol}
\begin{lstlisting}
# \textbf{square}
# Original loop

a, b = 0, 0
while true
  a = a + 2b + 1
  b = b + 1
end
\end{lstlisting}
\end{minipage}
\begin{minipage}[t]{\secondcol}
\begin{lstlisting}
# \textbf{square}
# \absynth{} loop
# Solver: Yices
a, b = 0, 0
while true
  a = a - 2b + 1
  b = b - 1
end
\end{lstlisting}
\end{minipage}
\begin{minipage}[t]{\thirdcol}
\begin{lstlisting}
# \textbf{square}
# \absynth{} loop
# Solver: Z3
a, b = 1/16, -1/4
while true
  a = a + 2b + 1
  b = b + 1
end
\end{lstlisting}
\end{minipage}

\end{center}

\captionspace
\caption{Example \texttt{square} with loop invariant \texttt{a == b\^{}2}}
\label{fig:experiments:b}
\figspace
\end{subfigure}
\begin{subfigure}{\textwidth}
\begin{center}
\begin{minipage}[t]{\firstcol}
\begin{lstlisting}
# \textbf{sum1}
# Original loop

a, b, c = 0, 0, 1
while true
  a = a + 1
  b = b + c
  c = c + 2
end
\end{lstlisting}
\end{minipage}
\begin{minipage}[t]{\secondcol}
\begin{lstlisting}
# \textbf{sum1}
# \absynth{} loop
# Solver: Yices
a, b, c = 1/2, 1/4, 2
while true
  a = a - 1/2
  b = b - 1/2 c + 3/4
  c = c - 1
end
\end{lstlisting}
\end{minipage}
\begin{minipage}[t]{\thirdcol}
\begin{lstlisting}
# \textbf{sum1}
# \absynth{} loop
# Solver: Z3
a, b, c = -5/8, 25/64, -1/4
while true
  a = a + 1
  b = b + c
  c = c + 2
end
\end{lstlisting}
\end{minipage}

\end{center}

\captionspace
\caption{Example \texttt{sum1} with input \texttt{1+2a == c \&\& 4b == (c-1)\^{}2}}
\label{fig:experiments:c}
\figspace
\end{subfigure}
\begin{subfigure}{\textwidth}
\begin{center}
\begin{minipage}[t]{\firstcol}
\begin{lstlisting}
# \textbf{intsqrt2}
# Original loop

y, r = 1/2 a0, 0
while true
    y = y - r
    r = r + 1
end
\end{lstlisting}
\end{minipage}
\begin{minipage}[t]{\secondcol}
\begin{lstlisting}
# \textbf{intsqrt2}
# \absynth{} loop
# Solver: Yices
y, r = 1/2 a0, 0
while true
  y = y + r - 1
  r = r - 1
end
\end{lstlisting}
\end{minipage}
\begin{minipage}[t]{\thirdcol}
\begin{lstlisting}
# \textbf{intsqrt2}
# \absynth{} loop
# Solver: Z3
y, r  = 1/2 a0 - 5/32, -1/4
while true
    y = y - r
    r = r + 1
end
\end{lstlisting}
\end{minipage}

\end{center}

\captionspace
\caption{Example \texttt{intsqrt2} with loop invariant \texttt{a0+r == r\^{}2+2y}}
\label{fig:experiments:d}
\figspace
\end{subfigure}
\begin{subfigure}{\textwidth}
\begin{center}
\begin{minipage}[t]{0.30\textwidth} 
\begin{lstlisting}
# \textbf{intcbrt}
# Original loop

x, r, s = a0, 1, 13/4
while true
  x = x - s
  s = s + 6r + 3
  r = r + 1
end
\end{lstlisting}
\end{minipage}
\begin{minipage}[t]{0.69\textwidth}
\begin{lstlisting}
# \textbf{intcbrt}
# \absynth{} loop
# Solver: Z3
x, s, r = 34/64 + a0, 7/16, -1/4
while true
  x = x - s
  s = s + 6r + 3
  r = r + 1
end
\end{lstlisting}
\end{minipage}

\end{center}

\captionspace
\caption{Example \texttt{intcbrt} with loop invariant \texttt{1+4a0+6r\^{}2==3r+4r\^{}3+4x \&\& 1/4+3r\^{}2==s}}
\label{fig:experiments:e}
\figspace
\end{subfigure}

\caption{Loops synthesized by \absynth{}, by reverse engineering examples from the state-of-the-art in polynomial invariant generation.}
\label{fig:synth:experiments}
\end{figure}

\revision{
\section{Beyond Loop Synthesis}\label{sec:beyond}

As argued in in Section~\ref{sec:intro}, our work has potential applications in other areas
related to synthesis. In this section we report on our experience  in using  to loop synthesis (i) in the context of deriving loops that are equivalent based on some metric and (ii) in the setting of 
teaching formal methods. 
 We note that former use case (i) has applications towards program/compiler optimization, as discussed below. 
Our experimental results from these use cases are summarized in Table~\ref{tab:experiments-new} and detailed next. Based on our initial experiments from Section~\ref{sec:experiments} showcasing the clear benefits of using \tool{Z3*} in \absynth{}, in this section, as well as in further applications of loop synthesis, we only deployed  \absynth{} with \tool{Z3} as an integrated backend solver.

\subsection{Synthesizing Equivalent Loops modulo Invariants}\label{sec:beyond:equiv}

Given a (potentially arbitrary) loop $L$  with loop variables $\vec{x}$, 
our work can be used to simplify $L$ while maintaining the polynomial invariants of $L$. That is, for a loop $L$ 
with a  polynomial relation $p(\vec{x})=0$ as its invariant, our approach can be used to synthesize a loop $L'$   such that $L'$ implements only affine operations over $\vec{x}$ and
$p(\vec{x})=0$ is an invariant of $L'$. In this case, we say that $L$ and $L'$ are \emph{equivalent modulo the invariant $p(\vec{x})=0$}. More generally, we define this form of equivalence between program loops as follows.


%

\begin{definition}[Loop Equivalence modulo Invariants]\label{def:equivModI}
Let ${L}$ and ${L'}$ be two program loops,  with   variable sets respectively denoted by $\vec{x}_{L}$ and $\vec{x}_{L'}$. Let $I(\vec{x}_{L})$ be an invariant property of $L$. We say that the   loops $L$ and $L'$ are \emph{equivalent modulo $I$} if  
$I(f(\vec{x}_{L}))$ is an invariant property of $L'$ for some bijective function $f \colon \vec{x}_{L} \to \vec{x}_{L'}$. 

We write  $L \equiv_{I} {L'}$ to denote that $L$ and $L'$ are equivalent modulo $I$. 

    
    %
\end{definition}

Note that we define loop equivalence modulo an invariant $I$  for arbitrary loops and arbitrary invariants. That is, Definition~\ref{def:equivModI} is not restricted to polynomial invariants, nor to program loops defined by our programming model~\eqref{eq:loop}. 

We next argue that our loop synthesis approach can be used to generate loops that are equivalent modulo an invariant $I$, as follows: 
\begin{enumerate}
    \item Given a loop $L$ with variables $\vec{x}_L$ and a polynomial relation $p(\vec{x}_L)=0$ as an invariant $I(\vec{x}_L)$ of $L$; 
    \item Synthesize a loop $L'$ with variables $\vec{x}_{L'}$  such that $I(\vec{x}_{L'})$ is an invariant of $L'$. 
\end{enumerate}
Note that in our approach to infer $L'$ as an equivalent loop of $L$ modulo $I$,  (i) we restrict our invariants $I$ to be polynomial relations among loop variables, but (ii) we do not impose our programming model constraints~\eqref{eq:loop} over  $L$. That is, our loop synthesis approach can be used to synthesize  loops $L'$ that are ``simpler" than an  arbitrary (not necessarily single-path affine) loop $L$.  For example,  while a  loop $L$ with invariant $I$ may  contain arbitrary polynomial assignments and nested conditionals, 
our loop synthesis approach can be used to derive a single-path loop $L'$ with only affine updates such that $L'$ is equivalent to $L$ modulo $I$. Example~\ref{ex:beyond:equiv} showcases such a use case of our loop synthesis approach: we simplify a multi-path loop into a single-path loop, while maintaining loop equivalence modulo an invariant. Such a simplification can be seen as an instance of program/compiler optimizations, for example in the generic setting of strength reduction, as argued in Section~\ref{sec:intro}. }

\begin{figure}
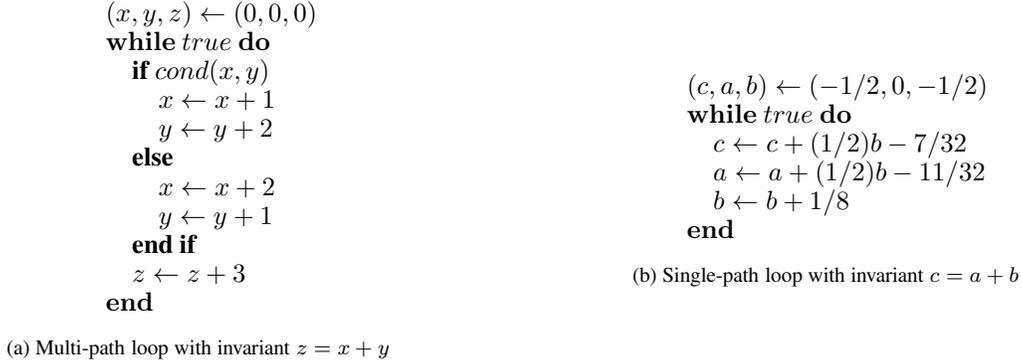

\captionsetup[sub]{justification=raggedright,singlelinecheck=false,margin=4em,skip=0pt}
\revision{
    \begin{subfigure}{.5\textwidth}
    \begin{center}
      \begin{tabular}{l}
      $(x, y, z) \gets (0, 0, 0)$\\
      $\WHILE~true~\DO$\\
      \quad$ \text{\bf if } cond(x,y)$\\
      \quad\quad $x \gets x+1$\\
      \quad\quad $y \gets y+2$\\
      \quad \text{\bf else }\\
      \quad\quad $x \gets x+2$\\
      \quad\quad $y \gets y+1$\\
     \quad \text{\bf end if }\\
     \quad $ z\gets z+3$\\ 
      $\END$\\
      {\color{white} add}
      \end{tabular}
    \end{center}
    \captionspace
    \caption{\revision{Multi-path loop with invariant $z=x+y$}}
    \label{fig:loop-with-if}
    \figspace
    \end{subfigure}
    \begin{subfigure}{.5\textwidth}
    \begin{center}
      \begin{tabular}{l}
      $(c, a, b) \gets (-1/2, 0, -1/2)$\\
      $\WHILE~true~\DO$\\
      \quad $c \gets c + (1/2)b - 7/32$\\
      \quad $a \gets a + (1/2)b - 11/32$\\
      \quad $b \gets b + 1/8$\\
      $\END$\\
      {\color{white} add}
      \end{tabular}
    \end{center}
    
    \captionspace
    \caption{\revision{Single-path loop with invariant $c = a + b$}}
    \label{fig:loop-without-if}
    \figspace
    \end{subfigure}
\caption{\revision{Examples of equivalent loops modulo invariants} }}
\label{fig:equivalent-loops}
\end{figure}

\revision{
\begin{example}\label{ex:beyond:equiv}
Consider the multi-path loop of Figure~\ref{fig:loop-with-if}, for which  $z - x - y = 0$ is an invariant. We write $I(x, y, z)$ to denote the polynomial invariant $z - x - y = 0$ of Figure~\ref{fig:loop-with-if}. We choose the bijective function $f = \{ (z, c), (x, a), (y, b) \}$ and use our loop synthesis approach to derive the single-path loop of Figure~\ref{fig:loop-without-if} from the invariant $I(f(x),f(y),f(z))$.   By soundness of our loop synthesis approach (Theorem~\ref{thm:sound-complete}), the invariant $I(f(x), f(y), f(z))$  denoting $c - a - b=0$ is an invariant of Figure~\ref{fig:loop-without-if}. Hence, the loops of  Figure~\ref{fig:loop-with-if} and Figure~\ref{fig:loop-without-if} are equivalent modulo ${z - x - y=0}$.
    
\end{example}

In Figure~\ref{fig:synth:experiments-unknown} we give further examples of  loops synthesized by our work such that the resulting loops are equivalent modulo an invariant. The examples in Figure~\ref{fig:synth:experiments-unknown} have been constructed by us to showcase the non-linear arithmetic reasoning features of \absynth. Further, the loops in Figure~\ref{fig:synth:experiments-unknown} are listed using the \absynth{} input languages, reporting further experiments made with \absynth{}. We note that our approach to loop equivalence exploits also the setting of parametrized loop synthesis (Section~\ref{sec:synth:param}): the initial values of loop variables of the respective equivalent loops may differ, as illustrated in the examples of Figure~\ref{fig:synth:experiments-unknown}. While our loop synthesis generates only affine loops of the form~\eqref{eq:loop}, Figure~\ref{fig:synth:experiments-unknown} illustrates that such loops can implement a wide range of affine updates over loop variables, including dependencies among updates,  while maintaining loop equivalence modulo an invariant. We finally note that, similarly to loop synthesis, loop equivalence modulo an invariant  can be used in \absynth{} with invariants representing arbitrary Boolean combinations of polynomial invariants; for example, conjunctive polynomial invariants as given in 
Figure~\ref{fig:synth:experiments-unknown}(a). 

}

\begin{table}
  \setlength{\arraycolsep}{3pt}
  \setlength{\tabcolsep}{2.4pt}
  \fontsize{8}{8}\selectfont
  \centering
  \def\arraystretch{.95}
  \revision{\begin{tabular}{@{}lrrrrrrrrrrrrrrrrrrrr@{}}
    \toprule
    \multirow{2}{*}{Instance} & \multirow{2}{*}{{\textsc{s}}} & \multirow{2}{*}{\textsc{i}} & \multirow{2}{*}{\textsc{d}} & \multirow{2}{*}{\textsc{c}} & \phantom{x} & \multicolumn{3}{c}{\tool{Z3*}} & \phantom{x} & \multicolumn{3}{c}{\tool{Z3* + Alg2}} \\ 
    \cmidrule{7-9} \cmidrule{11-13}
    &&&&&& \uni & \upper & \full && \uni & \upper & \full                                                           \\
    \midrule
    \texttt{cube\_conj}    & 5 & 2 & 3 & 90 && $11$ \UN & \TO \UN  & 
    \TO \UN && $174$ \UN & \TO \UN & \TO  \\
    \texttt{squared\_varied1}    & 4 & 1 & 2 & 47 && $13$ \UN & \TO \UN  & 
    \TO \UN && $1906$ \UN & \TO \UN & \TO  \\
    \texttt{square\_conj}    & 4 & 2 & 2 & 54 && $8$ \UN & $22$ \UN  & 
    \TO \UN && $2161$ \UN & $68$ \UN & \TO  \\
    \texttt{cube\_square}    & 4 & 1 & 3 & 60 && $10$ \UN & \TO \UN  & 
    \TO \UN && $2316$ \UN & \TO \UN & \TO  \\
    \texttt{sum\_of\_square}    & 5 & 1 & 2 & 92 && $46$ \UN & \TO \UN  & 
    \TO \UN && $65$ \UN & \TO \UN & \TO  \\
    \texttt{squared\_varied2}    & 4 & 1 & 2 & 50 && $31$ \UN & $69$ \UN  & 
    \TO \UN && $295$ \UN & $155$ \UN & \TO  \\
    \texttt{fmi1}      & 3 & 1 &  2 & 32 && $7$ \UN & $13$ \UN  & $22$        && $1619$ \UN & $21$       & $267$    \\
    \texttt{fmi2}      & 4 & 2 & 2 & 54 && $10$ \UN & $19$ \UN  & \TO        && $21$ \UN & $86$       & \TO    \\
    \texttt{fmi3}      & 4 & 2 & 2 & 55 && $12$       & $19$       & \TO        && $1840$ \UN & \TO \UN & \TO    \\
    \texttt{fmi4}    & 3 & 1 & 2 & 32 && $8$ \UN & $16$ \UN  & $22$ \UN && $13$ \UN & $34$ \UN & $84$  \\
    \texttt{fmi5}    & 3 & 1 & 2 & 32 && $7$ \UN & $14$ \UN  & 
    $19$ \UN && $7$ \UN & $217$ \UN & \TO  \\
    \bottomrule
  \end{tabular}
  \vskip1ex
  \def\arraystretch{1}
  \begin{tabular}{llcll}
    \textsc{s} & size of the recurrence system & &  & - timeout (60 seconds)\\
    \textsc{i} & number of polynomial invariants \\
    \textsc{d} & maximum monomial degree of constraints\\
    \textsc{c} & number of constraints\\
  \end{tabular}}
  \vspace{.5em}
\caption{\revision{\absynth{} results for using  loop synthesis for  (i)  the loop equivalence  benchmarks of  Figure~\ref{fig:synth:experiments:fm} and in  (ii) the teaching efforts of  formal methods from  Figure~\ref{fig:synth:experiments-unknown} (results in milliseconds)}}
  \label{tab:experiments-new}
\end{table}

\begin{figure}
  \captionsetup[sub]{justification=raggedright,singlelinecheck=false,margin=4em,skip=0pt}
  \begin{subfigure}{\textwidth}
\begin{center}

\begin{minipage}[t]{\firstcol}
\begin{lstlisting}[escapechar=\%]
# \textbf{cube\_conj}
# \absynth{} loop
# Solver: Z3
a, b, c, d = 0, 1/8, -1/16, 1/4
while true
    a = a + b - (1/2)d
    b = b - 1/2
    c = c + 2d - 1
    d = d - 1
end
\end{lstlisting}
\end{minipage}
\begin{minipage}[t]{\firstcol}
\begin{lstlisting}[escapechar=\%]
# \textbf{cube\_conj}
# \absynth{} loop
# Solver: Z3
a, b, c, d = 1/4, 1/8, -15/64, 1/2
while true
    a = a - b + (1/2)d - 1/8
    b = b - 1/4
    c = c + d - 1/4
    d = d - 1/2 
end
\end{lstlisting}
\end{minipage}

\end{center}

\captionspace
\caption{Example \texttt{cube\_conj} with loop invariant \texttt{a + 2b == d \&\& d\^{}2 + c == a\^{}3}}
\label{fig:experiments:unk1}
\figspace
\end{subfigure}
\begin{subfigure}{\textwidth}
\begin{center}
\begin{minipage}[t]{\firstcol}
\begin{lstlisting}
# \textbf{squared\_varied1}
# \absynth{} loop
# Solver: Z3
a, b, c = 7, 4, -1/2
while true
    a = -15/32 + a + (1/2)c
    b = -7/32 + b + (1/2)c
    c = 1/8 + c
end
\end{lstlisting}
\end{minipage}
\begin{minipage}[t]{\secondcol}
\begin{lstlisting}
# \textbf{squared\_varied1}
# \absynth{} loop
# Solver: Z3
b, a, c = 2, 9/2, 0
while true
    b = b - 1/4
    a = -15/32 + a + (1/2)c
    c = c + 1/8
end
\end{lstlisting}
\end{minipage}

\end{center}

\captionspace
\caption{Example \texttt{squared\_varied1} with loop invariant \texttt{a(b + 2c) == b\^{}2 + 5}}
\label{fig:experiments:unk2}
\figspace
\end{subfigure}
\begin{subfigure}{\textwidth}
\begin{center}
\begin{minipage}[t]{\firstcol}
\begin{lstlisting}
# \textbf{square\_conj}
# \absynth{} loop
# Solver: Z3
b, c, a = -1/4, 1/16, -1/4
while true
    b = -2/3 - (8/3)a + b
    c = 1 + 2a + c
    a = 1 + a
end
\end{lstlisting}
\end{minipage}
\begin{minipage}[t]{\secondcol}
\begin{lstlisting}
# \textbf{square\_conj}
# \absynth{} loop
# Solver: Z3
b, c, a = 0, 1/4, 1/2
while true
    b = 1/16 - (1/3)a  + b
    c = 1/64 + (1/4)a + c
    a = 1/8 + a
end
\end{lstlisting}
\end{minipage}

\end{center}

\captionspace
\caption{Example \texttt{square\_conj} with loop invariant \texttt{2a == 3b + 4c \&\& c == a\^{}2}}
\label{fig:experiments:unk3}
\figspace
\end{subfigure}
\begin{subfigure}{\textwidth}
\begin{center}
\begin{minipage}[t]{\firstcol}
\begin{lstlisting}
# \textbf{cube\_square}
# \absynth{} loop
# Solver: Z3
a, b, c = 31/512, -1/4, -1/8
while true
    a = a + 2b + 1
    b = b + 15/16 - (1/2)c
end
\end{lstlisting}
\end{minipage}
\begin{minipage}[t]{\secondcol}
\begin{lstlisting}
# \textbf{cube\_square}
# \absynth{} loop
# Solver: Z3
a, b, c = 1/8, 1/2, -1/2
while true
    a = 1/64 + a + (1/4)b
    b = -1/8 + b - (1/2)c
end
\end{lstlisting}
\end{minipage}

\end{center}

\captionspace
\caption{Example \texttt{cube\_square} with loop invariant \texttt{a == b\^{}2 + c\^{}3}}
\label{fig:experiments:unk4}
\figspace
\end{subfigure}
\begin{subfigure}{\textwidth}
\begin{center}
\begin{minipage}[t]{\firstcol} 
\begin{lstlisting}
# \textbf{sum\_of\_square}
# \absynth{} loop
# Solver: Z3
a, d, b, c = -1/8, 81/64, -1/2, 1
while true
    a = -5/4 + a - b + c
    d = -49/32 + d - (1/8)b + (13/8)c
    b = 1/8 + b
    c = 1/8 + c
end
\end{lstlisting}
\end{minipage}
\begin{minipage}[t]{\secondcol}
\begin{lstlisting}
# \textbf{sum\_of\_square}
# \absynth{} loop
# Solver: Z3
d, a, b, c = 9/8, -1, -1/4, 1/4
while true
    d = -1 - 1/4 a - (1/8)b + (17/4)c + d
    a = -1/4 + a + (1/2)b + 2c
    b = -1/2 + b
    c = 1/8 + c
end
\end{lstlisting}
\end{minipage}

\end{center}

\captionspace
\caption{Example \texttt{sum\_of\_square} with loop invariant \texttt{a\^{}2 + b\^{}2 + c\^{}2 == d}}
\label{fig:experiments:unk5}
\figspace
\end{subfigure}

\begin{subfigure}{\textwidth}
\begin{center}
\begin{minipage}[t]{\firstcol}
\begin{lstlisting}
# \textbf{squared\_varied2}
# \absynth{} loop
# Solver: Z3
a, b, c = 1/8, 1/2, 7/8
while true
    a = -31/68 + a - b + (4/17)c
    b = -1/2  + b
    c = -17/8 + c
end
\end{lstlisting}
\end{minipage}
\begin{minipage}[t]{\secondcol}
\begin{lstlisting}
# \textbf{squared\_varied2}
# \absynth{} loop
# Solver: Z3
a, b, c = 3/16, 1/8, 3/8
while true
    a = 1/2 + a - (1/2)b + (1/6)c
    b = b + 1/3
    c = c + 1
end
\end{lstlisting}
\end{minipage}

\end{center}

\captionspace
\caption{Example \texttt{squared\_varied2} with loop invariant \texttt{2a + 3b\^{}2 - ab == c + ab}}
\label{fig:experiments:unk6}
\figspace
\end{subfigure}

\caption{\revision{Examples of loops generated by \absynth{} that are equivalent modulo an invariant}}
\label{fig:synth:experiments-unknown}
\end{figure}

\revision{
\subsection{Loop Synthesis in Teaching Formal Methods}\label{sec:beyond:teach}

By further exploiting loop synthesis and its applications towards loop equivalence modulo invariant, 
we now detail our setting for using loop synthesis in automating some of our  efforts in teaching deductive verification at the TU Wien. Amid  online lecturing and examinations during the COVID-19 epidemic, coming up with best practices
to assess course performance was far from trivial. We faced this challenge, for example, during our ``Formal Methods in Computer Science -- FMI"  course in the  computer science master curriculum 
at TU Wien. In particular, we had to provide plausible solutions towards online examinations on topics of deductive verification, where students are asked to formally prove partial/total correctness assertions using Hoare logic. As such, our online exam sheets were designed by requiring solutions on (i) generating  loop invariants and variants, and (ii) using invariants and variants to conduct inductive proofs about program correctness. 
Yet, to minimize collaborative approaches towards solving online exam solutions, we also had to design  an online exam setting that
(i) provided individual exam sheets for each student enrolled in our course (about 350 enrolled students per study semester), while (ii) requiring  the same solving workload on each participant during  our online exam. By taking into account all these constraints, we used our loop synthesis approach to generate online exam problems on proving partial correctness, as follows: 
\begin{enumerate}

    \item We fixed a small set $\vec{x}$ of program variables, typically using at most $3$ variables; 
    
    \item We fixed a few polynomial equalities among $\vec{x}$ by considering polynomials of degree at most 3. In particular, we used at most $3$ such polynomial relations, that is  $p_1(\vec{x})=0$, $p_2(\vec{x})=0$, and $p_3(\vec{x})=0$, with each $p_i$ of degree at most~$3$; 
    
    \item We constructed a polynomial property $I(\vec{x})$ as a conjunctive formula among (some of) the fixed polynomial relations  $p_i(\vec{x})=0$;
    
    \item We used loop synthesis in \absynth{} to generate different loops $L$ that are equivalent modulo $I$. In the process of generating these loops, we only considered loops whose coefficients are given by relatively small integer/rational numbers; 
    
    \item We formulated exam problems by reverse engineering instances of our loop synthesis task, using the loops generated by \absynth. That is, our exam problems on proving partial correctness were generated using the following simplified template: 
    \emph{``
    Given the loop $L$, prove that $I$ is an inductive invariant of $L$"}, 
    where we set $L$ to be a loop generated by \absynth{} using the invariant $I$. 
\end{enumerate}

Figure~\ref{fig:synth:experiments:fm} showcases some loops generated by \absynth{} that have been used in the deductive verification exam problems of our FMI master course during 2020/2021. 

We note that, 
depending on the invariants  $I$ we deployed for loop synthesis, as well as on the loop bodies of the loops $L$ generated by \absynth{}, we also considered exam problems where: (i) we used simple linear inequalities as loop conditions bounding the number of loop iterations by a symbolic constant $N$ (as shown in our motivating examples from Figure~\ref{fig:Dafny}); (ii) stated pre-conditions $A$ and post-conditions $B$ on $L$; and (iii) asked  for proving partial correctness of the Hoare triple $\{A\}~L~\{B\}$, by requiring students to also generate the necessary invariants $I$. 

Based on the students' performances during our online exams, we believe that the exam problems generated by  \absynth{} were fair challenges for students, 
reducing collusion among them. Moreover, despite the fact that our FMI exams were conducted online since 2020, the overall FMI grade distributions on solved online exam sheets were very similar to previous, in-class examinations.

}

\begin{figure}
  \captionsetup[sub]{justification=raggedright,singlelinecheck=false,margin=4em,skip=0pt}
  \begin{subfigure}{\textwidth}
\begin{center}
\begin{minipage}[t]{\firstcol}
\begin{lstlisting}[escapechar=\%]
# \textbf{fmi1}
# Original loop

y, x = 0, 0
while true
    y = 3x + y
    x = x + 1
end
%~%
\end{lstlisting}
\end{minipage}
\begin{minipage}[t]{\secondcol}
\begin{lstlisting}[escapechar=\%]
# \textbf{fmi1}
# \absynth{} loop
# Solver: Z3
y, x = 15/32, -1/4
while true
    y = 3x + y
    x = x + 1
end
\end{lstlisting}
\end{minipage}
\begin{minipage}[t]{\thirdcol}
\begin{lstlisting}[escapechar=\%]
# \textbf{fmi1}
# \absynth{} loop
# Solver: Z3
y, x = -3/8, 1/2 
while true
    y = y + (3/8)x - 21/128
    x = x + 1/8
end
\end{lstlisting}
\end{minipage}
\end{center}

\captionspace
\caption{Example \texttt{fmi1} with loop invariant \texttt{2y == 3x(x - 1)}}
\label{fig:experiments:fm1}
\figspace
\end{subfigure}
\begin{subfigure}{\textwidth}
\begin{center}
\begin{minipage}[t]{\firstcol}
\begin{lstlisting}
# \textbf{fmi2}
# Original loop

x, y, z = 0, 0, 0
while true
    x = x + z + 1
    z = z + 2
    y = y + 1
end
\end{lstlisting}
\end{minipage}
\begin{minipage}[t]{\secondcol}
\begin{lstlisting}
# \textbf{fmi2}
# \absynth{} loop
# Solver: Z3
z, x, y = 1/4, 1/64, 1/8
while true
    z = z - 1
    x = x - y + 1/4
    y = y - 1/2
end
\end{lstlisting}
\end{minipage}
\begin{minipage}[t]{\thirdcol}
\begin{lstlisting}
# \textbf{fmi2}
# \absynth{} loop
# Solver: Z3
z, x, y = 1, 1/4, 1/2
while true
    z = 1/8 + z
    x = (1/8)y + x + 1/256
    y = y + 1/16
end
\end{lstlisting}
\end{minipage}

\end{center}

\captionspace
\caption{Example \texttt{fmi2} with loop invariant \texttt{z == 2y \&\& x == y\^{}2}}
\label{fig:experiments:fm2}
\figspace
\end{subfigure}
\begin{subfigure}{\textwidth}
\begin{center}
\begin{minipage}[t]{\firstcol}
\begin{lstlisting}
# \textbf{fmi3}
# Original loop

x, y, z = 0, 0, 1
while true
    x = x + 2
    y = y + 6x
    z = z + 1
end
\end{lstlisting}
\end{minipage}
\begin{minipage}[t]{\secondcol}
\begin{lstlisting}
# \textbf{fmi3}
# \absynth{} loop
# Solver: Z3
y, x, z = 27/32, -9/4, -1/8
while true
    y = (-1/4)x + y - 1/2 + (25/2)z
    x = x + 2
    z = 1 + z
end
\end{lstlisting}
\end{minipage}
\begin{minipage}[t]{\thirdcol}
\begin{lstlisting}
# \textbf{fmi3}
# \absynth{} loop
# Solver: Z3
y, x, z = 9/2, -3, -1/2
while true
    y = y + (1/2)x + 11/32 + (1/2)z
    x = x + 1/4
    z = 1/8 + z
end
\end{lstlisting}
\end{minipage}

\end{center}

\captionspace
\caption{Example \texttt{fmi3} with loop invariant \texttt{y == 3xz \&\& x == 2(z - 1)}}
\label{fig:experiments:fm3}
\figspace
\end{subfigure}
\begin{subfigure}{\textwidth}
\begin{center}
\begin{minipage}[t]{\firstcol}
\begin{lstlisting}
# \textbf{fmi4}
# Original loop

x, y = 2, 1
while true
    x = x + 4y + 2
    y = y + 1
end
\end{lstlisting}
\end{minipage}
\begin{minipage}[t]{\secondcol}
\begin{lstlisting}
# \textbf{fmi4}
# \absynth{} loop
# Solver: Z3
x, y = 1/8, -1/4
while true
    x = x + 4y + 2
    y = y + 1
end
\end{lstlisting}
\end{minipage}
\begin{minipage}[t]{\thirdcol}
\begin{lstlisting}
# \textbf{fmi4}
# \absynth{} loop
# Solver: Z3
x, y = 1/2, 1/2
while true
    x = (1/2)y + x + 1/32
    y = y + 1/8
end
\end{lstlisting}
\end{minipage}

\end{center}

\captionspace
\caption{Example \texttt{fmi4} with loop invariant \texttt{x == 2y\^{}2}}
\label{fig:experiments:fm4}
\figspace
\end{subfigure}
%
    
    
    

\begin{subfigure}{\textwidth}
\begin{center}
\begin{minipage}[t]{\firstcol}
\begin{lstlisting}
# \textbf{fmi5}
# Original loop

x, y = 0, 0
while true
    x = x + 1
    y = y - 10x + 5
end
\end{lstlisting}
\end{minipage}
\begin{minipage}[t]{\secondcol}
\begin{lstlisting}
# \textbf{fmi5}
# \absynth{} loop
# Solver: Z3
y, x = -5/16, -1/4
while true
    y = -10x + y - 5
    x = x + 1
end
\end{lstlisting}
\end{minipage}
\begin{minipage}[t]{\thirdcol}
\begin{lstlisting}
# \textbf{fmi5}
# \absynth{} loop
# Solver: Z3
y, x = -5/4, 1/2
while true
    y = y - (5/4)x - 5/64
    x = x + 1/8
end
\end{lstlisting}
\end{minipage}

\end{center}

\captionspace
\caption{Example \texttt{fmi5} with loop invariant \texttt{y + 5x\^{}2 == 0}}
\label{fig:experiments:fm5}
\figspace
\end{subfigure}

\caption{\revision{Loops synthesized by \absynth{} for online exam problems within the "Formal Methods in Computer Science -- FMI" master course at TU Wien.}}
\label{fig:synth:experiments:fm}
\end{figure}


\section{Related Work}\label{sec:related}
We now compare our loop synthesis work in relation to the state-of-the-art in program synthesis and algebraic reasoning about program loops. 
%

\paragraph{Synthesis.}
To the best of our knowledge, existing synthesis approaches are restricted to
linear invariants, see for instance~\cite{GulwaniPOPL10}, whereas our work supports loop
synthesis from non-linear polynomial properties. 
Existing approaches to SyGuS-based
synthesis~\cite{Alur18} 
differ in the (a) logical expressiveness of input
specifications,
(b) restrictions of the solutions space of programs to be
synthesized and (c) search strategies for solutions during synthesis.
The key difference between our work and the
state-of-the-art in synthesis is three-fold: (i) we consider logical
specifications as (non-linear) polynomial invariant relations, (ii)
synthesize loops with non-linear behaviour, and (iii) 
precisely characterize the set of all such loops to be synthesized. To the
best of our knowledge, existing approaches are restricted to linear
invariants, see e.g.~\cite{GulwaniPOPL10}, and no other approach can
automatically synthesize loops from
polynomial invariants. 
In what
follows, we discuss approaches that are mostly related to our work.
focusing only on works handling loops and/or
recursion. 

Counterexample-guided
synthesis~(CEGIS)~\cite{Alur18,Solar09,DilligPLDI18,SolarICML19}
refines 
the
deductive synthesis approach of~\cite{MannaW80} and restricts the search space of
programs by using templates and/or sketches~\cite{Solar09}. CEGIS provides an automated
approach for synthesizing programs $P$ from a given specification
$S$~\cite{Alur18,Solar09,SolarICML19}. For doing so, CEGIS uses 
input-output examples satisfying a specification $S$ to synthesize a candidate program $P$
that is consistent with the given inputs. Correctness
of the candidate program $P$ with respect to $S$ is then checked using
verification approaches, in particular using SMT-based reasoning. If
verification fails, a counterexample is generated as an input to $P$ that
violates $S$. This counterexample is then used in conjunction with the previous
set of input-outputs to revise synthesis and generate a new candidate program
$P$.
CEGIS-based approaches have been extended with machine/active
learning to control the selection of (counter-) examples satisfying $S$ and guide
the synthesis of $P$~\cite{DilligPLDI18}.
Unlike these methods, input
specifications to our approach are relational (invariant) properties describing
all, potentially infinite input-output examples of interest. Hence, we do not
rely on interactive refinement of our input but work with a precise
characterization of the set of input-output values of the program to be
synthesized. Similarly to sketches~\cite{Solar09,SolarICML19}, we consider loop templates restricting the
search for solutions to synthesis. Yet, our templates support non-linear
arithmetic (and hence multiplication), which is not yet the case
in~\cite{SolarICML19,DilligPLDI18}. We rely on fixed point reasoning and
properties of algebraic recurrences to encode the synthesis problem as a
constraint solving task in the theory of non-linear real arithmetic. 
We precisely characterize the set of all programs
satisfying our input specification, and as such, our approach does not exploit
learning to refine program candidates. On the other hand, our programming model
is more restricted than~\cite{SolarICML19,DilligPLDI18} in various aspects:
we only handle simple loops and only consider numeric data types and operations.

The programming by example approach of~\cite{GulwaniPOPL11} learns programs from
input-output examples and relies on lightweight interaction to refine the
specification of programs to be specified. The approach has further been
extended in~\cite{GulwaniICLR18} with machine learning, allowing to learn
programs from just one (or even none) input-output example by using a simple
supervised learning setup. Program synthesis from input-output examples is shown
to be successful for recursive programs~\cite{GulwaniCAV13}, yet synthesizing
loops and handling non-linear arithmetic is not yet supported by this
line of research.
Our work does not learn programs from
observed input-output examples, but uses loop invariants to fully characterize
the intended behavior of the program to be synthesized. We precisely
characterize the solution space of loops to be synthesized by a system of
algebraic recurrences, without using statistical models supporting
machine learning. Thanks to its precision, our method is however computationally
 more expensive and does not yet scale to large programs/data sets.

A related approach to our work is given in~\cite{KuncakEMSOFT13}, where a fixed-point implementation for an approximated
real-valued polynomial specification is presented. For doing so, the
method of~\cite{KuncakEMSOFT13} uses
genetic programming~\cite{GP08} to guide the search
for synthesized programs and combines heuristic search with
with abstract
interpretation~\cite{CousotC77}
to estimate and refine the (floating-point) error
bound of the inferred fixed-point implementation. While the underlying abstract
interpreter is precise for linear expressions, precision of the synthesis is lost
for non-linear arithmetic. Unlike~\cite{KuncakEMSOFT13},
we consider polynomial specification in the abstract algebra of real-closed
fields and do not address challenges rising from machine
reals. 
As a consequence, we precisely synthesize loops from
polynomial invariants, by inherently relying on non-linear reasoning. 

\paragraph{Algebraic Reasoning.} Compared to works on invariant generation~\cite{Rodriguez-CarbonellK07,HumenbergerJK17,KincaidCBR18,Worrell18},
the only common aspect between these works and our synthesis
method is the use of linear recurrences to capture the  functional
behavior of program loops. Yet, our work is conceptually different
from~\cite{Rodriguez-CarbonellK07,HumenbergerJK17,KincaidCBR18,Worrell18},
as 
we reverse engineer invariant generation and do not rely on the ideal structure/Zariski closure of
polynomial invariants. We do not use ideal theory nor Gr\"obner bases
computation to generate invariants from loops; rather, we generate
loops from invariants by formulating and solving PCPs. 


\section{Conclusions}
\label{sec:conclusion}

We proposed a syntax-guided synthesis procedure for synthesizing loops
over affine assignments from polynomial invariants. We consider loop templates and use reasoning
over recurrence equations modeling the loop behavior. The key ingredient of our
work comes with translating the loop synthesis problem into a polynomial
constraint problem and showing that this constraint problem precisely captures
all solutions to the loop synthesis problem. Additional heuristics for
solving our constraints have been also implemented  in our new tool
\absynth{} for loop synthesis. 

Directions for future work include a complexity analysis of our algorithm;
further investigating the properties of our constraint problems for improving
the scalability of our procedure; generalizing our approach to multi-path loops
and inequality invariants; restricting the solution space to integers or bounded
domains; extending \absynth{} with reasoning support for arbitrary algebraic
numbers; and  understanding and encoding the best optimization measures for loop
synthesis in the context of program/compiler optimization
approaches.



\subsection*{Acknowledgments} We thank Sumit Gulwani and Manuel Kauers
for valuable discussions on ideas leading to this work. We acknowledge
funding from the ERC Consolidator Grant ARTIST 101002685, the ERC Starting Grant SYMCAR 639270, the ERC Proof of Concept Grant SYMELS 842066, the Wallenberg Academy
Fellowship TheProSE, the FWF research
project LogiCS  W1255-N23, and the WWTF grant ProbInG ICT19-018.

\bibliographystyle{alpha}
\bibliography{references}

\end{document}